\UseRawInputEncoding
\documentclass[superscriptaddress, twocolumn, amsmath, amssymb, aps,prl, letterdraft, notitlepage,longbibliography]{revtex4-1}
\usepackage{graphicx,graphics,epsfig,subfigure,times,bm,bbm,amssymb,amsmath,amsfonts,amsthm,mathrsfs,MnSymbol}
\usepackage[matrix,frame,arrow]{xypic}
\usepackage[normalem]{ulem}
\usepackage{slashed}
\usepackage{color}
\usepackage[usenames,dvipsnames,svgnames,table]{xcolor}
\usepackage[english]{babel}
\usepackage{verbatim}
\usepackage{tabularx}
\definecolor{darkblue}{rgb}{0.0,0.0,0.3}
\usepackage[colorlinks=true,
            linkcolor=red,
            urlcolor= darkblue,
            citecolor=blue]{hyperref}

\newcommand{\bea}{\begin{eqnarray}}
\newcommand{\eea}{\end{eqnarray}}

\begin{document}
\title{Quantum Nondemolition Photon Counting With a Hybrid Electromechanical Probe}

\author{Junjie Liu}
\address{Department of Chemistry and Centre for Quantum Information and Quantum Control,
University of Toronto, 80 Saint George St., Toronto, Ontario, M5S 3H6, Canada}
\author{Hsing-Ta Chen}
\address{Department of Chemistry, University of Pennsylvania, Philadelphia, Pennsylvania 19104, USA}
\author{Dvira Segal}
\address{Department of Chemistry and Centre for Quantum Information and Quantum Control,
University of Toronto, 80 Saint George St., Toronto, Ontario, M5S 3H6, Canada}
\address{Department of Physics, 60 Saint George St., University of Toronto, Toronto, Ontario, M5S 1A7, Canada}

\begin{abstract}
Quantum nondemolition (QND) measurements of photons is a much pursued endeavor in the field of quantum optics and quantum information processing. 
Here we propose a novel hybrid optoelectromechanical platform that integrates a cavity system with a hybrid electromechanical probe for QND photon counting. 
Building upon a mechanical-mode-mediated nonperturbative electro-optical dispersive coupling, our protocol performs the QND photon counting measurement by means of the current-voltage characteristics of the probe. 
In particular, we show that the peak voltage shift of the differential conductance is linearly dependent on the photon occupation number, thus providing a sensitive measure of the photon number, especially in the strong optomechanical coupling regime. 
Given that our proposed hybrid system is compatible with state-of-the-art experimental techniques, we discuss its implementations and anticipate applications in quantum optics and polariton physics. 
\end{abstract}

\date{\today}

\maketitle

\paragraph{Introduction.--}
Over the past decades, quantum non-demolition (QND)
measurements \cite{Caves.80.RMP,Braginsky.96.RMP,Bocko.96.RMP,Raimond.01.RMP,Wiseman.09.NULL} have been implemented in many experiments that require ultimate sensitivity, such as gravitational-wave detection \cite{Braginsky547}, with a growing recent interests in quantum information processing and storage applications \cite{lupascu_quantum_2007}.
Particularly, in the quantum optics community, it is well appreciated that a QND measurement scheme can evade the adverse effect of quantum back-action during the measurement \cite{Caves.80.RMP,Braginsky.96.RMP,Bocko.96.RMP}, such that an experiment observable can be repeatedly measured without perturbing the underlying quantum state.
Therefore, exploiting the theory of QND measurements, researchers have demonstrated novel strategies for probing quantum characteristics, including quantum fluctuation sensing \cite{PhysRevLett.97.133601,PhysRevA.90.043848,Appel10960} and photon counting \cite{Holland.91.PRL,Friberg.92.PRL,Jacobs.94.PRA,Brune.90.PRL,Nogues.99.N,Grangier.98.N,PhysRevA.66.063814,Guerlin.07.N,Schuster.07.N,Haroche.PS.09,Johnson.10.NP,Ludwig.12.PRL,Reiserer.S.13,Peaudecerf.14.PRL,Besse.18.PRX,Kono.18.NP,Grimsmo.20.A}, pushing the limit of quantum technology.

Conditions for an ideal QND measurement were formulated by Imoto et~al. \cite{Imoto.85.PRA}. As far as a general quantum measurement is concerned, one usually considers a signal observable ${O}_s$ of the measured system (with the unperturbed Hamiltonian ${H}_s$) and a readout observable ${O}_p$ of the probe system (with the unperturbed Hamiltonian ${H}_p$) coupled through an interaction operator ${H}_I$.
Following the strict definition of an ideal QND-type measurement \cite{Imoto.85.PRA}, ${O}_s$ and ${O}_p$ are a QND pair if the following mathematical conditions are satisfied: (i) ${H}_I={H}_I({O}_s)$, (ii) $[{H}_I,{O}_s]=0$, (iii) $[{H}_I,{O}_p]\neq0$, and (iv) $[{H}_s,{O}_s] =0$.
In the context of photon number measurements, to achieve QND-type detection of photons---namely to avoid any optical absorption of photons during the measurement---one common choice is to probe the photon of the measured system (${O}_s={a}_s^\dagger{a}_s$ as the system photon number operator) by another optical mode (${O}_p={a}_p-{a}_p^\dagger$ where ${a}_p$ is the annihilation operator of the probe photon) with a non-linear interaction Hamiltonian (e.g. in a Kerr medium ${H}_I\propto{a}_s^\dagger{a}_s{a}_p^\dagger{a}_p$) using optical interferometers \cite{Grangier.98.N,PhysRevA.66.063814,Imoto.85.PRA,Munro.05.PRA}. 
Nevertheless, such measurement schemes are almost always restricted to certain medium-dependent frequencies at which strong non-linearity can emerge. 

Recent advances in quantum optics have enabled QND photon number measurements that do not rely on the material non-linearity through strong light--matter interactions using cavity or circuit quantum electrodynamical systems \cite{Holland.91.PRL,Nogues.99.N,Guerlin.07.N,Schuster.07.N,Haroche.PS.09,Johnson.10.NP,Ludwig.12.PRL,Reiserer.S.13,Peaudecerf.14.PRL,Besse.18.PRX,Kono.18.NP,Malz.20.PRR}.
In these QND schemes, cavity photons are coupled to a probe atomic system (e.g. a two-level system) with large atom-cavity detuning so that the cavity photon frequency is off-resonant with the electronic transition of the atom. One key point for such QND measurements is that, in the large detuning limit, the Stark shift of the atomic transition as induced by the atom-photon coupling is approximately linear-dependent on the photon number \cite{Blais.04.PRA}, so that the dispersive phase shift of the atom can be measured and served as a QND readout of the photon number. 
That being said, such a readout is based on a perturbative treatment (one neglects the higher-order terms to the interaction Hamiltonian in the large detuning limit), and it is known that the higher-order corrections will inevitably cause the measurement to demolish the measured quantum state, thereby fundamentally restricting the applicability of a perturbative dispersive readout \cite{PhysRevA.77.060305,Slichter.12.PRL,Sank.16.PRL}. In practice, such measurement-induced demolition leads to a progressive damage to the qubit and cavity states as one aims to continuously monitor the qubit \cite{Blais.04.PRA,Slichter.12.PRL}. 
Furthermore, in the large photon occupation scenario, it costs significant resource of atoms to reach the desired accuracy of measured photon numbers \cite{Grangier.98.N,Guerlin.07.N,Peaudecerf.14.PRL}.

With this background in mind, it is clear that to achieve an ideal QND photon measurement, one needs a \emph{non-perturbative} dispersive coupling that does not rely on the linear dispersive limit \cite{Dassonneville.20.PRX}.
%
\begin{figure}[tbh!]
 \centering
\includegraphics[width=0.75\columnwidth] {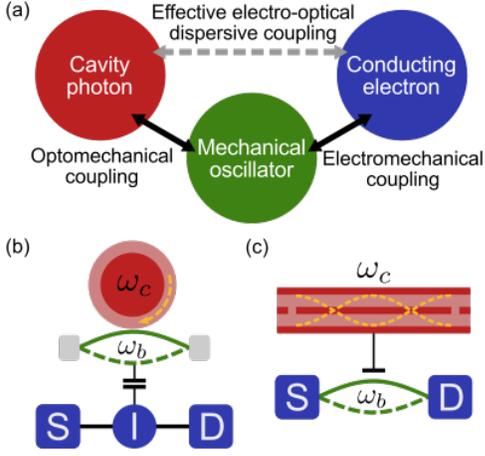} 
\caption{(a) Schematic of an opto-electromechanical system 
supporting an effective electro-optic dispersive coupling. 
Direct interactions are marked by solid double arrows. 
Possible realizations are depicted in panels (b) and (c).
In panel (b), the electromechanical probe consists of a 
nanomechanical oscillator with frequency $\omega_b$ capacitively coupled to a solid-state single electron transistor comprising a source (S), a drain (D) and an island (I),
see Refs. \cite{Knobel.03.N,LaHaye.S.04,Mozyrsky.04.PRL,Blencowe.04.PR}. 
An optical microtoroid cavity \cite{Aspelmeyer.14.RMP} supporting a whispering 
gallery mode of frequency $\omega_c$ is coupled to the mechanical oscillator 
by the near field effect. 
In panel (c), the electromechanical probe is realized by a voltage-biased carbon nanotube 
\cite{Steele.S.09,Wen.20.NP}, which is further coupled to a microwave cavity, 
see Ref. \cite{Blien.20.NC}.
} 
\label{fig:fig1}
\end{figure}
%
In this letter, we propose an alternative yet novel QND photon counting measurement scheme 
combining optomechanics \cite{Aspelmeyer.14.RMP} and nano electromechanics \cite{Wen.20.NP},
which holds promise to circumvent aforementioned limitations of existing schemes. Particularly, we suggest to use a hybrid opto-electromechanical protocol that integrates the measured cavity with an electromechanical probe; see Fig.~\ref{fig:fig1} for an illustration. Cavity photons interact with the mechanical mode of the probe through the radiation-pressure force \cite{Aspelmeyer.14.RMP} which imprints the photon occupation in the mechanical motion. A single electron transistor (SET), being the electronic component of the probe, is exploited to measure the mechanical motion to which it is capacitively coupled. 
Most importantly, we demonstrate that this hybrid opto-electromechanical system yields a mechanical-mode-mediated \emph{non-perturbative} electro-optical dispersive coupling (see Fig. \ref{fig:fig1} (a)), enabling a single-shot QND readout of photon number via charge current measurements of the SET. 
In comparison, our proposal outperforms existing QND photon counting schemes \cite{Grangier.98.N,PhysRevA.66.063814,Imoto.85.PRA,Munro.05.PRA,Holland.91.PRL,Nogues.99.N,Guerlin.07.N,Schuster.07.N,Haroche.PS.09,Johnson.10.NP,Ludwig.12.PRL,Reiserer.S.13,Peaudecerf.14.PRL,Besse.18.PRX,Kono.18.NP,Malz.20.PRR} 
in several ways: 
(i) An electronic signal is more reliable and robust  against environmental noise
than phase shift readout used with atom probes. 
(ii) In our scheme the resource cost is independent of the photon numbers, 
in sharp contrast to schemes using atomic meters,
see, e.g., Refs. \cite{Guerlin.07.N,Peaudecerf.14.PRL}. 
(iii) Our scheme avoids progressive damage to the photon state, which arises in other systems 
due to the nonideality of the QND scheme.

To make the applicability of our protocol
clear, we discuss its implementations with currently-available experimental 
conditions for optomechanical \cite{Naik.06.N,Anetsberger.09.NP,Groblacher.09.N,Clerk.10.RMP,Stannigel.10.PRL,Teufel.11.N,Teufel.11.Na,Verhagen.12.N,Aspelmeyer.14.RMP,Xiang.13.RMP,Blien.20.NC} and electromechanical systems \cite{Knobel.03.N,LaHaye.S.04,Mozyrsky.04.PRL,Steele.S.09,Lassagne.S.09,Lee.10.PRL,Winger.11.OP,Yeo.14.NN,Okazaki.16.NC,Wen.20.NP},
see Fig. \ref{fig:fig1} (b) and (c). 
In the strong optomechanical coupling regime, we show that the voltage shift of 
the differential conductance peak provides a sensitive measure for 
the number of photons stored in the cavity.

\paragraph{Hybrid opto-electromechanical system.--}
We consider an opto-electromechanical system \cite{Midolo.18.NN} $H$ which includes a high-quality cavity with a single photon mode ($a$, $a^{\dagger}$) to be measured, and the electromechanical probe which consists of a mechanical mode ($b$, $b^{\dagger}$), the SET conductor ($d$, $d^{\dagger}$) and the source ($S$) and drain ($D$) electrodes ($c_{kv}$, $c_{kv}^\dagger$ for $v=S,D$).
Here we denote each component by their annihilation and creation operators respectively. 
The total Hamiltonian $H$ reads (setting $\hbar=1$, $e=1$, $k_B=1$ and Fermi energy $\epsilon_F=0$ hereafter)
\bea\label{h0}
H &=& \epsilon_0d^{\dagger}d+\omega_c a^{\dagger}a+\omega_bb^{\dagger}b \nonumber\\
&&-g_0a^{\dagger}a(b^{\dagger}+b)+\lambda d^{\dagger}d(b^{\dagger}+b) \nonumber\\
&&+\sum_{k,v=S,D}\Big[\epsilon_{kv}c_{kv}^{\dagger}c_{kv}+t_{kv}(c_{kv}^{\dagger}d+d^{\dagger}c_{kv})\Big].
\eea
where $\omega_b$ and $\omega_c$ are the frequency of the mechanical mode and the cavity photon, respectively. 
$g_0$ is the single-photon optomechanical coupling strength \cite{Aspelmeyer.14.RMP}. 
$\lambda$ denotes the electromechanical coupling strength. Here the SET is assumed to be in the sequential tunneling regime such that its conductor can be described by a single electronic level at electrostatic energy level $\epsilon_0$, coupled to a collection of electrons in the two electrodes with energies $\epsilon_{kv}$, $v=S,D$.
This coupling is characterized by the spectral density defined as $\Gamma_v(\epsilon)=\pi\sum_kt_{kv}^2\delta(\epsilon-\epsilon_{kv})$. Throughout the study, we consider the wide-band limit, $\Gamma_v(\epsilon)=\Gamma_v$ \cite{Wingreen.89.PRB}.
For solid-state SET, we focus here on normal conductors 
\footnote{We point out that Eq.~\eqref{h0} is applicable in the sequential tunneling regime regardless of whether the solid-state SET is superconducting or not \cite{Clerk.05.NJP,Rodrigues.05.NJP}. The only difference is that in the case of an superconducting SET, one should interpret $d^{\dagger}d$ as the occupation operator for quasi-particles. 
For our case, although we are in the superconducting regime ($T\sim 100$ mK), the normal-state description is still applicable as one can apply an out-of-plane magnetic field to turn a superconducting SET to a normal one \cite{Knobel.03.N}.}.

We further include a dissipation Hamiltonian, $H_{\mathrm{tot}}=H+H_{\mathrm{diss}}$, where
$H_{\mathrm{diss}}$ represents the damping of the mechanical mode by its thermal environment at an ambient temperature $T_0$; this dissipation term will be treated at the level of an input-output theory \cite{SM.20.NULL}.  Here we do not include the cavity mode damping with the understanding that typical QND measurements are performed within a time scale faster than that of such decay process \cite{Guerlin.07.N}. While, in general, the mechanical damping occurs on a time scale that is much slower that that of cavity photon decay process, we keep it since it plays a crucial role for determining the current-voltage characteristics of the SET \cite{SM.20.NULL}.

To reveal that the mechanical mode mediates an electro-optical dispersive coupling, we introduce a unitary transformation
\begin{equation}
\mathcal{G}~=~\exp\Big[-g_0(b^{\dagger}-b)a^{\dagger}a/\omega_b\Big]\otimes\exp\Big[\lambda(b^{\dagger}-b)d^{\dagger}d/\omega_b\Big].
\end{equation}
The transformed system Hamiltonian $\tilde{H}\equiv\mathcal{G}H\mathcal{G}^{\dagger}$ reads
\bea\label{eq:tilde_h}
\tilde{H} &=& \left(\epsilon_0-\frac{\lambda^2}{\omega_b}\right)d^{\dagger}d+\omega_c a^{\dagger}a+\omega_bb^{\dagger}b+H_K+H_I\nonumber\\
&&+\sum_{k,v=S,D}\Big[\epsilon_{kv}c_{kv}^{\dagger}c_{kv}+t_{kv}(c_{kv}^{\dagger}\tilde{d}+\tilde{d}^{\dagger}c_{kv})\Big].
\eea
Here $\tilde{d}\equiv\mathcal{D}_{\lambda}^{\dagger}d$ with a displacement operator $\mathcal{D}_{\lambda}\equiv\exp[(b^{\dagger}-b)\lambda/\omega_b]$. It should be noted that we account for the full radiation-pressure coupling, rather than its linearized form \cite{Aspelmeyer.14.RMP}. The effect of this transformation on $H_{\mathrm{diss}}$ is negligible as the coupling between the mechanical resonator and thermal environment is typically weak \cite{SM.20.NULL,Aspelmeyer.14.RMP}.

The transformed Hamiltonian Eq.~\eqref{eq:tilde_h} yields an electro-optical dispersive coupling $H_I=\frac{2\lambda g_0}{\omega_b}a^{\dagger}ad^{\dagger}d$.  
We emphasize that this interaction is \emph{nonperturbative}, namely no Hamiltonian truncation is involved \cite{Blais.04.PRA,Dassonneville.20.PRX}.
Generally, $H_I$ can be interpreted as either an electron-number-dependent shift of the cavity frequency or, vice versa, a photon-number-dependent shift of the electronic level.
The transformation also generates an effective Kerr non-linear term $H_K=-\frac{g_0^2}{\omega_b}(a^{\dagger}a)^2$ between the cavity photons.
However, $H_K$ conserves the cavity photon number and hence has no impact on the QND photon counting.

Next, since we neglected cavity losses during the measurement process, the photon occupation $\bar{n}_p\equiv\langle a^{\dagger}a\rangle$ can be considered a time-independent observable. 
Therefore, for a given photon number $\bar{n}_p$, the dispersive coupling leads to a renormalized electrostatic energy of the SET conductor
\begin{equation}\label{eq:tilde_epsilon}
\tilde{\epsilon}_n~\equiv~\epsilon_0-\frac{\lambda^2}{\omega_b}+2\frac{\lambda g_0}{\omega_b}\bar{n}_p.
\end{equation}
Notably, this renormalized electronic level $\tilde{\epsilon}_n$ sets the condition for resonant electron transport. Thus, a shift in the electronic energy leads to a linear shift of the bias voltage at which resonant transport occurs.

\paragraph{Current-voltage characteristics of the SET.--}
The steady state charge current of the SET serves to probe the photon number. Based on a generalized input-output method \cite{Liu.20.PRB,Liu.20.PRBa,SM.20.NULL}, we arrive at the 
expression for the current out of the source
\begin{equation}\label{eq:JL}
\langle J_S\rangle~=~\frac{4\Gamma_S\Gamma_D}{\Gamma}\int\,\frac{d\epsilon}{2\pi}\mathcal{G}(\epsilon)[n_F^S(\epsilon)-n_F^D(\epsilon)].
\end{equation}
Here, $\Gamma=\Gamma_S+\Gamma_D$ and $n_F^v(\epsilon)=[\exp[(\epsilon-\mu_v)/T_0]+1]^{-1}$ is the Fermi-Dirac distribution of the $v$th lead with $\mu_v$ the chemical potential and $T_0$ the ambient temperature. Here, for simplicity, we assume a symmetric bias drop for the SET, namely, $\mu_S=-\mu_D=V/2$ with $V$ the voltage bias. The generalized transmission function reads
\begin{equation}
\mathcal{G}(\epsilon)~\equiv~\mathrm{Re}\Big[\int_{0}^{\infty}d\tau e^{-(\Gamma+i\tilde{\epsilon}_n-i\epsilon)\tau}B_{\lambda}(\tau)\Big].
\end{equation}
Here, `Re' takes the real part, $\tilde{\epsilon}_n$ is given by Eq.~\eqref{eq:tilde_epsilon} and $B_{\lambda}(\tau)$ denotes a mechanical mode correlation function \cite{SM.20.NULL}
\bea\label{eq:bb_corr}
B_{\lambda}(\tau) &=& \exp\Big[-\frac{\lambda^2}{\omega_b^2}\int \frac{d\omega}{\pi}\frac{\gamma_{\mathrm{eff}}}{\gamma_{\mathrm{eff}}^2+(\omega-\omega_{b})^2}\nonumber\\
&&\times\Big(\mathrm{coth}(\omega/2T_{\mathrm{eff}})(1-\cos\omega\tau)+i\sin\omega\tau\Big)\Big].
\eea

To arrive at Eqs.~(\ref{eq:JL})--(\ref{eq:bb_corr}), we employ an effective bath description \cite{Armour.04.PRB,Clerk.04.PRB,Clerk.05.NJP,Rodrigues.05.NJP} to capture the dynamics of the mechanical mode. Based on the time scale separation between the fast electron dynamics and the slow mechanical motion, this treatment allows us to include both the intrinsic thermal dissipation of the mechanical mode (with damping rate $\gamma_0$ and temperature $T_0$) and the back-action from the conducting electrons (as approximated in terms of an extra thermalized bath with damping rate $\gamma_1$ and temperature $T_1$, which is proportional to the applied voltage bias $V$) \cite{SM.20.NULL,Armour.04.PRB,Clerk.05.NJP,Rodrigues.05.NJP}. Consequently, the overall effect of the thermal dissipation and the back-action on the mechanical mode is described by an effective damping rate $\gamma_{\mathrm{eff}}\equiv\gamma_0+\gamma_1$ from an effective thermal bath characterized by an effective temperature $T_{\mathrm{eff}}=(\gamma_0T_0+\gamma_1T_1)/\gamma_{\mathrm{eff}}$ \cite{Clerk.04.PRB,Armour.04.PRB,Clerk.05.NJP,Naik.06.N}. Typically, $\gamma_1/\gamma_0\sim20-50$  \cite{Naik.06.N,Bennett.10.PRL}.

\paragraph{QND photon counting scheme.--}
We are now ready to formulate a QND photon counting measurement protocol using the current-voltage characteristics of the SET as a readout observable. 
First, we easily verify that, given the dispersive coupling $H_I=\frac{2\lambda g_0}{\omega_b}n_pd^{\dagger}d$, the cavity photon number operator $n_p\equiv a^{\dagger}a$ and the charge current operator $J_S\equiv i\sum_kt_{kS}(c_{kS}^{\dagger}\tilde{d}-\tilde{d}^{\dagger}c_{kS})$ satisfy the conditions of QND measurements \cite{Imoto.85.PRA}, i.e. (i) $H_I=H_I(n_p)$, (ii) $[H_I,n_p]=0$, and (iii) $[H_I,J_S]\neq0$.
Therefore, one can infer the photon number without disturbing the cavity field from a charge current measurement.

In practice, we propose the following two-step protocol to determine the photon number confined in the cavity by contrasting the following measurements:
(1) By coupling an empty cavity to the electromechanical probe, we determine the peak position of the differential conductance $\partial\langle J_S\rangle/\partial V$ of the SET (denoted as the reference voltage $V_0^{\ast}$). The peak position corresponds to the island energy $\epsilon_0-\frac{\lambda^2}{\omega_b}$. 
(2) Pumping a photonic field with a finite yet unknown photon occupation $\bar{n}_p$ to the empty cavity \cite{Peaudecerf.14.PRL}, the peak of the differential conductance will shift and appear at voltage bias $V_n^{\ast}$ corresponding to the renormalized island energy $\tilde{\epsilon}_n$. 

Following this protocol, the photon occupation of the cavity can be simply inferred from the voltage difference $V_n^{\ast}-V_0^{\ast}$,
\begin{equation}\label{eq:n_measure}
\bar{n}_{p,\mathrm{measure}}~=~\frac{V_n^{\ast}-V_0^{\ast}}{2\Delta \epsilon}.
\end{equation}
This constitutes one of main results of our work. We obtain this expression assuming a symmetric bias drop for the SET, with the understanding that our scheme is not limited to this scenario \cite{SM.20.NULL}.
The resolution of the photon number measurement is determined by $\Delta \epsilon\equiv2\lambda g_0/\omega_b$. 
Such a QND photon counting can reach high sensibility by increasing either the optomechanical coupling strength ($g_0$) or the relative electromechanical counterpart ($\lambda/\omega_b$), or ideally, both.
Although here we rely on the differential conductance for the measurement of the voltage shift $V_n^{\ast}-V_0^{\ast}$, one can also resort to the second-order derivative $\partial^2\langle J_S\rangle/\partial V^2$ and identify the voltage values $V_n^{\ast}$ from its node, which also marks the onset of resonant transport.

\paragraph{Experimental feasibility.--}
We now discuss the feasibility of the proposed hybrid platform with state-of-the-art nanoscale fabrication technologies for quantum cavity optomechanical systems \cite{Naik.06.N,Anetsberger.09.NP,Groblacher.09.N,Clerk.10.RMP,Stannigel.10.PRL,Teufel.11.N,Teufel.11.Na,Verhagen.12.N,Aspelmeyer.14.RMP,Xiang.13.RMP,Blien.20.NC} and electromechanical counterparts \cite{Knobel.03.N,LaHaye.S.04,Mozyrsky.04.PRL,Steele.S.09,Lassagne.S.09,Lee.10.PRL,Winger.11.OP,Yeo.14.NN,Okazaki.16.NC,Wen.20.NP,Khivrich.19.NN}. 
Electromechanical systems integrating a solid-state SET have been fabricated in Refs. \cite{Knobel.03.N,LaHaye.S.04,Mozyrsky.04.PRL,Blencowe.04.PR}. A straightforward implementation of our proposed hybrid system would be to couple this SET-oscillator system to an optical microresonator \cite{Aspelmeyer.14.RMP} 
by optical near field effects \cite{Anetsberger.09.NP}, see Fig.~\ref{fig:fig1} (b). 
The emerging field of carbon nanotube optomechanics \cite{Blien.20.NC} 
provides another promising platform for realizing the proposed hybrid system,
with the carbon nanotube further playing the role of an electromechanical probe,
 see Fig.~\ref{fig:fig1} (c). For demonstration purposes, 
below we focus on the former, Fig.~\ref{fig:fig1} (b)  setup,
as this field is more mature.  

First, we justify the assumption that $\bar{n}_p$ remains time-independent during the charge current measurement. For a typical solid-state SET, the electrostatic capacitance is $C_{\Sigma}\sim 400$ aF \cite{Devoret.00.N,LaHaye.S.04} and the total junction resistance is $R\sim100$~k$\Omega$, so that the electron tunneling time can be estimated by $\tau_e=2RC_{\Sigma}\approx0.1$ns \cite{Armour.04.PRB}.
We consider a high-quality cavity in which the cavity photon decays at a damping rate ($\kappa/2\pi$) of the order of MHz \cite{Groblacher.09.N,Teufel.11.N,Teufel.11.Na,Verhagen.12.N} and the lifetime is about $1/\kappa\approx1~\mu$s.

Second, as far as the peak voltage shift is concerned, the photon number resolution measured by Eq.~(\ref{eq:n_measure}) is determined by $\Delta \epsilon=2\lambda g_0/\omega_b$.
Here we choose the electromechanical coupling to be weak (typically $\lambda/\omega_b\sim 0.1$ \cite{Ouyang.09.PRB}) so that the effective bath description is valid \cite{Rodrigues.05.NJP,Clerk.05.NJP}. So far, the achieved single-photon optomechanical coupling strength $g_0/2\pi$ is ranging from few Hz to hundreds of kHz \cite{Aspelmeyer.14.RMP}.  

Finally, notwithstanding, one legitimate concern is that zero-point quantum fluctuation and weak optomechanical coupling \cite{Aspelmeyer.14.RMP} may make this QND measurement ineffective at a single-photon level. 
Indeed, we focus here on experiments carried out with a large photon number (at least on the order of ${\bar{n}_p}\approx 10^6$) for achieving a strong optomechanical coupling (see, e.g., Ref. \cite{Verhagen.12.N}), i.e. $g_0\sqrt{\bar{n}_p}>\kappa$.
In this multi-photon scenario, our QND measurement protocol should yield a peak voltage shift of the order $\bar{n}_p\Delta \epsilon \sim\Gamma=\tau_e^{-1}\approx 10^{-6}$ eV with peak broadening determined by $\Gamma$ and the effective temperature $T_{\mathrm{eff}}$.
Hence, under these experimental conditions, the effect of zero-point quantum fluctuation should not prevent us to observe a clear differential conductance peak shift.

\paragraph{Proof-of-principle simulation.--}
To access the efficacy of the proposed QND photon counting scheme in the strong optomechanical coupling regime, we provide a proof-of-principle simulation of Eq.~\eqref{eq:n_measure} for given photon occupation $\bar{n}_p$ using Eqs.~(\ref{eq:JL})--(\ref{eq:bb_corr}). 
%
\begin{figure}[tbh!]
 \centering
\includegraphics[width=0.9\columnwidth] {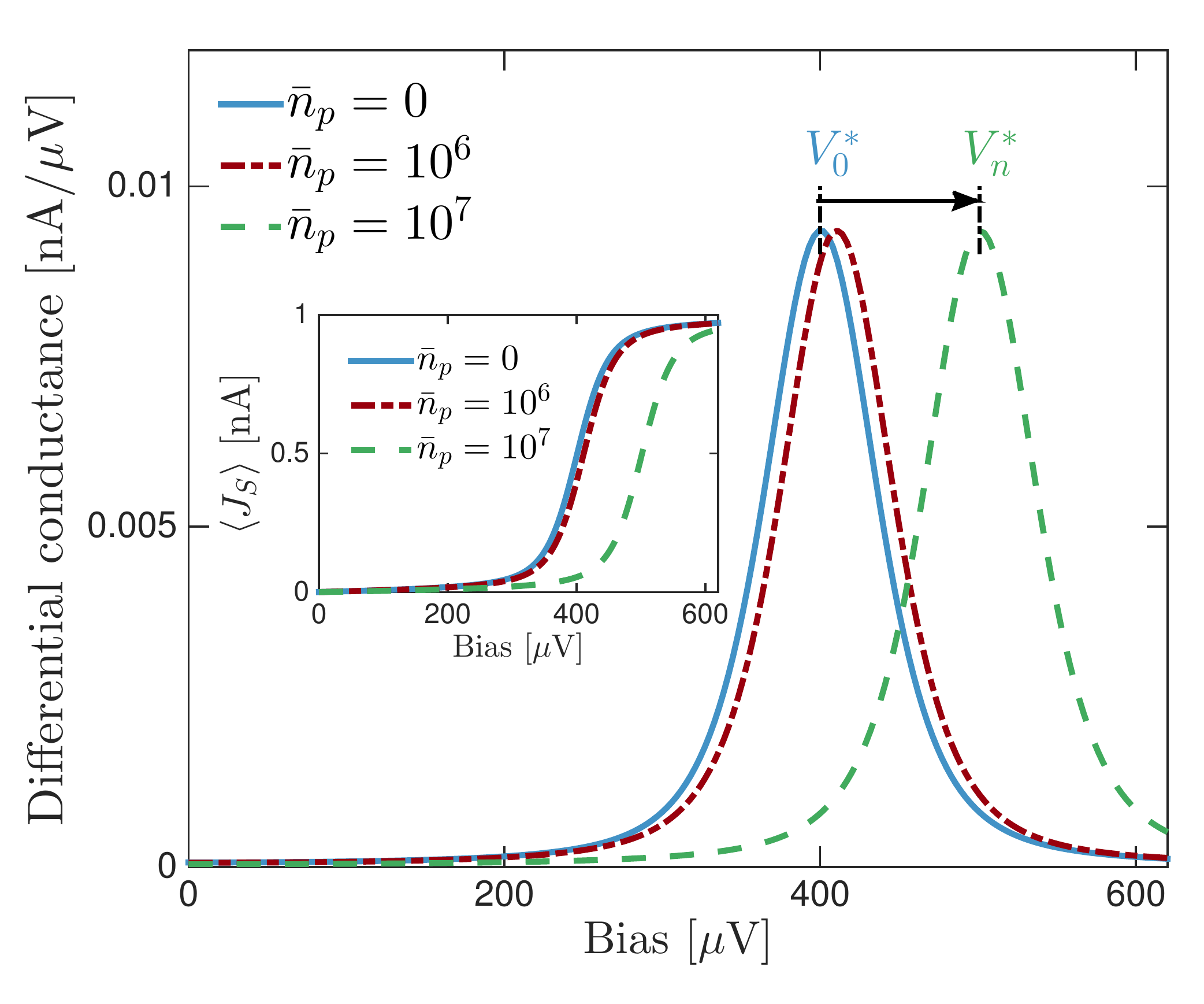} 
\caption{Differential conductance of the SET for $\bar{n}_p=0$ (blue solid line), $\bar{n}_p=1\times10^6$ (red dash-dotted line) and $\bar{n}_p=1\times10^7$ (green dashed line). A peak shift from marked $V_0^{\ast}$ to $V_n^{\ast}$ is clearly visible. The inset depicts the current-voltage characteristics for $\bar{n}_p=0$ (blue solid line), $\bar{n}_p=1\times10^6$ (red dash-dotted line) and $\bar{n}_p=1\times10^7$ (green dashed line). Parameters are $\mu_S=-\mu_D=V/2$, $T_0=100$ mK, $g_0/2\pi=3$ kHz, $\omega_b/2\pi=10$ MHz, $\gamma_0/2\pi=2$ kHz, $\gamma_1/\gamma_0=30$, $T_1=V/4$, $C_{\Sigma}=400$ aF such that $\epsilon_0=1/(2C_{\Sigma})\simeq200$ $\mu$eV, $R=100$ k$\Omega$, $\lambda/\omega_b=0.2$ and  $\Gamma_{S,D}=\hbar/(2RC_{\Sigma})\sim 4.1$ $\mu$eV.} 
\label{fig:fig2}
\end{figure}
%
Fig. \ref{fig:fig2} depicts the differential conductance of the SET using experimentally available parameters. We clearly observe a voltage shift of the differential conductance peak compared with that obtained when the cavity is empty, namely, $\bar{n}_p=0$, thereby demonstrating the feasibility of the proposed QND photon counting scheme under current experimental conditions. 

Although intriguing, there are few remarks that are worth mentioning: (i) We have adopted the relation $T_1=V/4$ \cite{Armour.04.PRB,Clerk.05.NJP}, see definitions below Eq. (\ref{eq:bb_corr}). Strictly speaking, this expression is valid for large voltage bias of the order of $\epsilon_0$ \cite{Armour.04.PRB,Clerk.05.NJP}. Hence, quantitatively, our calculation overestimates the back-action from the conducting electrons to the mechanical mode at low voltages. However, this overestimation should have a negligible effect on demonstrating our QND protocol as we are interested in the resonance peak of the differential conductance rather than its broadening. (ii) The proposed QND photon counting scheme should be equally applicable in superconducting SET with modified electron tunneling rates \cite{Clerk.05.NJP}. Hence we expect that the voltage shift due to a finite photon occupation can still serve as an accessible QND measure in the superconducting case.

\paragraph{Conclusion.--}
We proposed an experimentally feasible QND measurement for cavity photon counting using an electromechanical probe.  
Our scheme builds upon a mechanical-mode-mediated coupling between cavity photons and conducting electrons, enabling a single-shot QND readout of photon number through the measurement of the SET charge current.
We further demonstrated the feasibility of the measurement protocol by simulating the current-voltage characteristics of the SET in a strong optomechanical coupling regime achieved with a large photon occupation.   

Looking forward, while reaching sufficiently strong single-photon optomechanical coupling is still challenging today \cite{Aspelmeyer.14.RMP,Liu.18.CPB}, we expect that a single-photon QND measurement with the proposed hybrid scheme can be realized in the near future exploiting experimental advances in 
cold-atom  \cite{Brennecke.S.08,Murch.08.NP}, photon-crystal \cite{Davanco.12.OP}, microwave \cite{Teufel.11.N,Teufel.11.Na,Zoepfl.20.PRL}, and carbon nanotube research \cite{Ares.16.PRL,Blien.20.NC}. These optomechanical systems have already showed great promise to reach the strong single-photon coupling regime.
With the ability to measure few photons, we should be able to non-destructively identify different cavity photon statistics, such as Poisson and Bose-Einstein distributions, by repeating the QND measurement and depicting the photon number histogram. 
Furthermore, if we include molecular systems within the cavity, this QND measurement may provide a direct probe to investigate polariton excitations (hybrid light-matter excitation when the molecular system is strongly coupled to cavity photons), rather than relying on far-field photon emission \cite{frisk_kockum_ultrastrong_2019}. 
Lastly, while we have conveniently neglected time-dependence of the photon occupation in the present paper, extension of this QND scheme for observing multi-photon correlation functions should reveal more quantum properties, such as photon bunching and antibunching \cite{press2007photon}, representing an exciting new direction for quantum optics.

\paragraph{Acknowledgement.--}
The authors thank Prof. Abraham Nitzan for insightful discussions and constructive comments. J. Liu and D. Segal acknowledge support from the Natural Sciences and Engineering Research Council (NSERC) of Canada Discovery Grant and the Canada Research Chairs Program. H.-T. Chen thanks the U.S. Department of Energy, Office of Science, Office of Basic Energy Sciences (Award Number DE-SC0019397).


\begin{thebibliography}{79}%
\makeatletter
\providecommand \@ifxundefined [1]{%
 \@ifx{#1\undefined}
}%
\providecommand \@ifnum [1]{%
 \ifnum #1\expandafter \@firstoftwo
 \else \expandafter \@secondoftwo
 \fi
}%
\providecommand \@ifx [1]{%
 \ifx #1\expandafter \@firstoftwo
 \else \expandafter \@secondoftwo
 \fi
}%
\providecommand \natexlab [1]{#1}%
\providecommand \enquote  [1]{``#1''}%
\providecommand \bibnamefont  [1]{#1}%
\providecommand \bibfnamefont [1]{#1}%
\providecommand \citenamefont [1]{#1}%
\providecommand \href@noop [0]{\@secondoftwo}%
\providecommand \href [0]{\begingroup \@sanitize@url \@href}%
\providecommand \@href[1]{\@@startlink{#1}\@@href}%
\providecommand \@@href[1]{\endgroup#1\@@endlink}%
\providecommand \@sanitize@url [0]{\catcode `\\12\catcode `\$12\catcode
  `\&12\catcode `\#12\catcode `\^12\catcode `\_12\catcode `\%12\relax}%
\providecommand \@@startlink[1]{}%
\providecommand \@@endlink[0]{}%
\providecommand \url  [0]{\begingroup\@sanitize@url \@url }%
\providecommand \@url [1]{\endgroup\@href {#1}{\urlprefix }}%
\providecommand \urlprefix  [0]{URL }%
\providecommand \Eprint [0]{\href }%
\providecommand \doibase [0]{http://dx.doi.org/}%
\providecommand \selectlanguage [0]{\@gobble}%
\providecommand \bibinfo  [0]{\@secondoftwo}%
\providecommand \bibfield  [0]{\@secondoftwo}%
\providecommand \translation [1]{[#1]}%
\providecommand \BibitemOpen [0]{}%
\providecommand \bibitemStop [0]{}%
\providecommand \bibitemNoStop [0]{.\EOS\space}%
\providecommand \EOS [0]{\spacefactor3000\relax}%
\providecommand \BibitemShut  [1]{\csname bibitem#1\endcsname}%
\let\auto@bib@innerbib\@empty
\bibitem [{\citenamefont {Caves}\ \emph {et~al.}(1980)\citenamefont {Caves},
  \citenamefont {Thorne}, \citenamefont {Drever}, \citenamefont {Sandberg},\
  and\ \citenamefont {Zimmermann}}]{Caves.80.RMP}%
  \BibitemOpen
  \bibfield  {author} {\bibinfo {author} {\bibfnamefont {C.~M.}\ \bibnamefont
  {Caves}}, \bibinfo {author} {\bibfnamefont {K.~S.}\ \bibnamefont {Thorne}},
  \bibinfo {author} {\bibfnamefont {R.~W.~P.}\ \bibnamefont {Drever}}, \bibinfo
  {author} {\bibfnamefont {V.~D.}\ \bibnamefont {Sandberg}}, \ and\ \bibinfo
  {author} {\bibfnamefont {M.}~\bibnamefont {Zimmermann}},\ }\bibfield  {title}
  {\enquote {\bibinfo {title} {On the measurement of a weak classical force
  coupled to a quantum-mechanical oscillator. i. issues of principle},}\ }\href
  {\doibase 10.1103/RevModPhys.52.341} {\bibfield  {journal} {\bibinfo
  {journal} {Rev. Mod. Phys.}\ }\textbf {\bibinfo {volume} {52}},\ \bibinfo
  {pages} {341--392} (\bibinfo {year} {1980})}\BibitemShut {NoStop}%
\bibitem [{\citenamefont {Braginsky}\ and\ \citenamefont
  {Khalili}(1996)}]{Braginsky.96.RMP}%
  \BibitemOpen
  \bibfield  {author} {\bibinfo {author} {\bibfnamefont {V.~B.}\ \bibnamefont
  {Braginsky}}\ and\ \bibinfo {author} {\bibfnamefont {F.~Ya.}\ \bibnamefont
  {Khalili}},\ }\bibfield  {title} {\enquote {\bibinfo {title} {Quantum
  nondemolition measurements: the route from toys to tools},}\ }\href {\doibase
  10.1103/RevModPhys.68.1} {\bibfield  {journal} {\bibinfo  {journal} {Rev.
  Mod. Phys.}\ }\textbf {\bibinfo {volume} {68}},\ \bibinfo {pages} {1--11}
  (\bibinfo {year} {1996})}\BibitemShut {NoStop}%
\bibitem [{\citenamefont {Bocko}\ and\ \citenamefont
  {Onofrio}(1996)}]{Bocko.96.RMP}%
  \BibitemOpen
  \bibfield  {author} {\bibinfo {author} {\bibfnamefont {M.~F.}\ \bibnamefont
  {Bocko}}\ and\ \bibinfo {author} {\bibfnamefont {R.}~\bibnamefont
  {Onofrio}},\ }\bibfield  {title} {\enquote {\bibinfo {title} {On the
  measurement of a weak classical force coupled to a harmonic oscillator:
  experimental progress},}\ }\href {\doibase 10.1103/RevModPhys.68.755}
  {\bibfield  {journal} {\bibinfo  {journal} {Rev. Mod. Phys.}\ }\textbf
  {\bibinfo {volume} {68}},\ \bibinfo {pages} {755--799} (\bibinfo {year}
  {1996})}\BibitemShut {NoStop}%
\bibitem [{\citenamefont {Raimond}\ \emph {et~al.}(2001)\citenamefont
  {Raimond}, \citenamefont {Brune},\ and\ \citenamefont
  {Haroche}}]{Raimond.01.RMP}%
  \BibitemOpen
  \bibfield  {author} {\bibinfo {author} {\bibfnamefont {J.~M.}\ \bibnamefont
  {Raimond}}, \bibinfo {author} {\bibfnamefont {M.}~\bibnamefont {Brune}}, \
  and\ \bibinfo {author} {\bibfnamefont {S.}~\bibnamefont {Haroche}},\
  }\bibfield  {title} {\enquote {\bibinfo {title} {Manipulating quantum
  entanglement with atoms and photons in a cavity},}\ }\href {\doibase
  10.1103/RevModPhys.73.565} {\bibfield  {journal} {\bibinfo  {journal} {Rev.
  Mod. Phys.}\ }\textbf {\bibinfo {volume} {73}},\ \bibinfo {pages} {565--582}
  (\bibinfo {year} {2001})}\BibitemShut {NoStop}%
\bibitem [{\citenamefont {Wiseman}\ and\ \citenamefont
  {Milburn}(2009)}]{Wiseman.09.NULL}%
  \BibitemOpen
  \bibfield  {author} {\bibinfo {author} {\bibfnamefont {H.~M.}\ \bibnamefont
  {Wiseman}}\ and\ \bibinfo {author} {\bibfnamefont {G.~J.}\ \bibnamefont
  {Milburn}},\ }\href@noop {} {\emph {\bibinfo {title} {Quantum Measurement and
  Control}}}\ (\bibinfo  {publisher} {Cambridge University Press, Cambridge},\
  \bibinfo {year} {2009})\BibitemShut {NoStop}%
\bibitem [{\citenamefont {Braginsky}\ \emph {et~al.}(1980)\citenamefont
  {Braginsky}, \citenamefont {Vorontsov},\ and\ \citenamefont
  {Thorne}}]{Braginsky547}%
  \BibitemOpen
  \bibfield  {author} {\bibinfo {author} {\bibfnamefont {V.~B.}\ \bibnamefont
  {Braginsky}}, \bibinfo {author} {\bibfnamefont {Y.~I.}\ \bibnamefont
  {Vorontsov}}, \ and\ \bibinfo {author} {\bibfnamefont {K.}~\bibnamefont
  {Thorne}},\ }\bibfield  {title} {\enquote {\bibinfo {title} {Quantum
  nondemolition measurements},}\ }\href {\doibase 10.1126/science.209.4456.547}
  {\bibfield  {journal} {\bibinfo  {journal} {Science}\ }\textbf {\bibinfo
  {volume} {209}},\ \bibinfo {pages} {547--557} (\bibinfo {year}
  {1980})}\BibitemShut {NoStop}%
\bibitem [{\citenamefont {Lupa{\c s}cu}\ \emph {et~al.}(2007)\citenamefont
  {Lupa{\c s}cu}, \citenamefont {Saito}, \citenamefont {Picot}, \citenamefont
  {de~Groot}, \citenamefont {Harmans},\ and\ \citenamefont
  {Mooij}}]{lupascu_quantum_2007}%
  \BibitemOpen
  \bibfield  {author} {\bibinfo {author} {\bibfnamefont {A.}~\bibnamefont
  {Lupa{\c s}cu}}, \bibinfo {author} {\bibfnamefont {S.}~\bibnamefont {Saito}},
  \bibinfo {author} {\bibfnamefont {T.}~\bibnamefont {Picot}}, \bibinfo
  {author} {\bibfnamefont {P.~C.}\ \bibnamefont {de~Groot}}, \bibinfo {author}
  {\bibfnamefont {C.~J. P.~M.}\ \bibnamefont {Harmans}}, \ and\ \bibinfo
  {author} {\bibfnamefont {J.~E.}\ \bibnamefont {Mooij}},\ }\bibfield  {title}
  {\enquote {\bibinfo {title} {Quantum non-demolition measurement of a
  superconducting two-level system},}\ }\href {\doibase 10.1038/nphys509}
  {\bibfield  {journal} {\bibinfo  {journal} {Nat. Phys.}\ }\textbf {\bibinfo
  {volume} {3}},\ \bibinfo {pages} {119--123} (\bibinfo {year}
  {2007})}\BibitemShut {NoStop}%
\bibitem [{\citenamefont {Arcizet}\ \emph {et~al.}(2006)\citenamefont
  {Arcizet}, \citenamefont {Cohadon}, \citenamefont {Briant}, \citenamefont
  {Pinard}, \citenamefont {Heidmann}, \citenamefont {Mackowski}, \citenamefont
  {Michel}, \citenamefont {Pinard}, \citenamefont
  {Fran\ifmmode~\mbox{\c{c}}\else \c{c}\fi{}ais},\ and\ \citenamefont
  {Rousseau}}]{PhysRevLett.97.133601}%
  \BibitemOpen
  \bibfield  {author} {\bibinfo {author} {\bibfnamefont {O.}~\bibnamefont
  {Arcizet}}, \bibinfo {author} {\bibfnamefont {P.-F.}\ \bibnamefont
  {Cohadon}}, \bibinfo {author} {\bibfnamefont {T.}~\bibnamefont {Briant}},
  \bibinfo {author} {\bibfnamefont {M.}~\bibnamefont {Pinard}}, \bibinfo
  {author} {\bibfnamefont {A.}~\bibnamefont {Heidmann}}, \bibinfo {author}
  {\bibfnamefont {J.-M.}\ \bibnamefont {Mackowski}}, \bibinfo {author}
  {\bibfnamefont {C.}~\bibnamefont {Michel}}, \bibinfo {author} {\bibfnamefont
  {L.}~\bibnamefont {Pinard}}, \bibinfo {author} {\bibfnamefont
  {O.}~\bibnamefont {Fran\ifmmode~\mbox{\c{c}}\else \c{c}\fi{}ais}}, \ and\
  \bibinfo {author} {\bibfnamefont {L.}~\bibnamefont {Rousseau}},\ }\bibfield
  {title} {\enquote {\bibinfo {title} {High-sensitivity optical monitoring of a
  micromechanical resonator with a quantum-limited optomechanical sensor},}\
  }\href {\doibase 10.1103/PhysRevLett.97.133601} {\bibfield  {journal}
  {\bibinfo  {journal} {Phys. Rev. Lett.}\ }\textbf {\bibinfo {volume} {97}},\
  \bibinfo {pages} {133601} (\bibinfo {year} {2006})}\BibitemShut {NoStop}%
\bibitem [{\citenamefont {Xu}\ and\ \citenamefont
  {Taylor}(2014)}]{PhysRevA.90.043848}%
  \BibitemOpen
  \bibfield  {author} {\bibinfo {author} {\bibfnamefont {X.}~\bibnamefont
  {Xu}}\ and\ \bibinfo {author} {\bibfnamefont {J.~M.}\ \bibnamefont
  {Taylor}},\ }\bibfield  {title} {\enquote {\bibinfo {title} {Squeezing in a
  coupled two-mode optomechanical system for force sensing below the standard
  quantum limit},}\ }\href {\doibase 10.1103/PhysRevA.90.043848} {\bibfield
  {journal} {\bibinfo  {journal} {Phys. Rev. A}\ }\textbf {\bibinfo {volume}
  {90}},\ \bibinfo {pages} {043848} (\bibinfo {year} {2014})}\BibitemShut
  {NoStop}%
\bibitem [{\citenamefont {Appel}\ \emph {et~al.}(2009)\citenamefont {Appel},
  \citenamefont {Windpassinger}, \citenamefont {Oblak}, \citenamefont {Hoff},
  \citenamefont {Kj{\ae}rgaard},\ and\ \citenamefont {Polzik}}]{Appel10960}%
  \BibitemOpen
  \bibfield  {author} {\bibinfo {author} {\bibfnamefont {J.}~\bibnamefont
  {Appel}}, \bibinfo {author} {\bibfnamefont {P.~J.}\ \bibnamefont
  {Windpassinger}}, \bibinfo {author} {\bibfnamefont {D.}~\bibnamefont
  {Oblak}}, \bibinfo {author} {\bibfnamefont {U.~B.}\ \bibnamefont {Hoff}},
  \bibinfo {author} {\bibfnamefont {N.}~\bibnamefont {Kj{\ae}rgaard}}, \ and\
  \bibinfo {author} {\bibfnamefont {E.~S.}\ \bibnamefont {Polzik}},\ }\bibfield
   {title} {\enquote {\bibinfo {title} {Mesoscopic atomic entanglement for
  precision measurements beyond the standard quantum limit},}\ }\href {\doibase
  10.1073/pnas.0901550106} {\bibfield  {journal} {\bibinfo  {journal} {Proc.
  Natl. Acad. Sci. U.S.A.}\ }\textbf {\bibinfo {volume} {106}},\ \bibinfo
  {pages} {10960--10965} (\bibinfo {year} {2009})}\BibitemShut {NoStop}%
\bibitem [{\citenamefont {Holland}\ \emph {et~al.}(1991)\citenamefont
  {Holland}, \citenamefont {Walls},\ and\ \citenamefont
  {Zoller}}]{Holland.91.PRL}%
  \BibitemOpen
  \bibfield  {author} {\bibinfo {author} {\bibfnamefont {M.~J.}\ \bibnamefont
  {Holland}}, \bibinfo {author} {\bibfnamefont {D.~F.}\ \bibnamefont {Walls}},
  \ and\ \bibinfo {author} {\bibfnamefont {P.}~\bibnamefont {Zoller}},\
  }\bibfield  {title} {\enquote {\bibinfo {title} {Quantum nondemolition
  measurements of photon number by atomic beam deflection},}\ }\href {\doibase
  10.1103/PhysRevLett.67.1716} {\bibfield  {journal} {\bibinfo  {journal}
  {Phys. Rev. Lett.}\ }\textbf {\bibinfo {volume} {67}},\ \bibinfo {pages}
  {1716--1719} (\bibinfo {year} {1991})}\BibitemShut {NoStop}%
\bibitem [{\citenamefont {Friberg}\ \emph {et~al.}(1992)\citenamefont
  {Friberg}, \citenamefont {Machida},\ and\ \citenamefont
  {Yamamoto}}]{Friberg.92.PRL}%
  \BibitemOpen
  \bibfield  {author} {\bibinfo {author} {\bibfnamefont {Stephen~R.}\
  \bibnamefont {Friberg}}, \bibinfo {author} {\bibfnamefont {Susumu}\
  \bibnamefont {Machida}}, \ and\ \bibinfo {author} {\bibfnamefont {Yoshihisa}\
  \bibnamefont {Yamamoto}},\ }\bibfield  {title} {\enquote {\bibinfo {title}
  {Quantum-nondemolition measurement of the photon number of an optical
  soliton},}\ }\href {\doibase 10.1103/PhysRevLett.69.3165} {\bibfield
  {journal} {\bibinfo  {journal} {Phys. Rev. Lett.}\ }\textbf {\bibinfo
  {volume} {69}},\ \bibinfo {pages} {3165--3168} (\bibinfo {year}
  {1992})}\BibitemShut {NoStop}%
\bibitem [{\citenamefont {Jacobs}\ \emph {et~al.}(1994)\citenamefont {Jacobs},
  \citenamefont {Tombesi}, \citenamefont {Collett},\ and\ \citenamefont
  {Walls}}]{Jacobs.94.PRA}%
  \BibitemOpen
  \bibfield  {author} {\bibinfo {author} {\bibfnamefont {K.}~\bibnamefont
  {Jacobs}}, \bibinfo {author} {\bibfnamefont {P.}~\bibnamefont {Tombesi}},
  \bibinfo {author} {\bibfnamefont {M.~J.}\ \bibnamefont {Collett}}, \ and\
  \bibinfo {author} {\bibfnamefont {D.~F.}\ \bibnamefont {Walls}},\ }\bibfield
  {title} {\enquote {\bibinfo {title} {Quantum-nondemolition measurement of
  photon number using radiation pressure},}\ }\href {\doibase
  10.1103/PhysRevA.49.1961} {\bibfield  {journal} {\bibinfo  {journal} {Phys.
  Rev. A}\ }\textbf {\bibinfo {volume} {49}},\ \bibinfo {pages} {1961--1966}
  (\bibinfo {year} {1994})}\BibitemShut {NoStop}%
\bibitem [{\citenamefont {Brune}\ \emph {et~al.}(1990)\citenamefont {Brune},
  \citenamefont {Haroche}, \citenamefont {Lefevre}, \citenamefont {Raimond},\
  and\ \citenamefont {Zagury}}]{Brune.90.PRL}%
  \BibitemOpen
  \bibfield  {author} {\bibinfo {author} {\bibfnamefont {M.}~\bibnamefont
  {Brune}}, \bibinfo {author} {\bibfnamefont {S.}~\bibnamefont {Haroche}},
  \bibinfo {author} {\bibfnamefont {V.}~\bibnamefont {Lefevre}}, \bibinfo
  {author} {\bibfnamefont {J.~M.}\ \bibnamefont {Raimond}}, \ and\ \bibinfo
  {author} {\bibfnamefont {N.}~\bibnamefont {Zagury}},\ }\bibfield  {title}
  {\enquote {\bibinfo {title} {Quantum nondemolition measurement of small
  photon numbers by rydberg-atom phase-sensitive detection},}\ }\href {\doibase
  10.1103/PhysRevLett.65.976} {\bibfield  {journal} {\bibinfo  {journal} {Phys.
  Rev. Lett.}\ }\textbf {\bibinfo {volume} {65}},\ \bibinfo {pages} {976--979}
  (\bibinfo {year} {1990})}\BibitemShut {NoStop}%
\bibitem [{\citenamefont {Nogues}\ \emph {et~al.}(1999)\citenamefont {Nogues},
  \citenamefont {Rauschenbeutel}, \citenamefont {Osnaghi}, \citenamefont
  {Brune}, \citenamefont {Raimond},\ and\ \citenamefont
  {Haroche}}]{Nogues.99.N}%
  \BibitemOpen
  \bibfield  {author} {\bibinfo {author} {\bibfnamefont {G.}~\bibnamefont
  {Nogues}}, \bibinfo {author} {\bibfnamefont {A.}~\bibnamefont
  {Rauschenbeutel}}, \bibinfo {author} {\bibfnamefont {S.}~\bibnamefont
  {Osnaghi}}, \bibinfo {author} {\bibfnamefont {M.}~\bibnamefont {Brune}},
  \bibinfo {author} {\bibfnamefont {J.~M.}\ \bibnamefont {Raimond}}, \ and\
  \bibinfo {author} {\bibfnamefont {S.}~\bibnamefont {Haroche}},\ }\bibfield
  {title} {\enquote {\bibinfo {title} {Seeing a single photon without
  destroying it},}\ }\href {https://doi.org/10.1038/22275} {\bibfield
  {journal} {\bibinfo  {journal} {Nature}\ }\textbf {\bibinfo {volume} {400}},\
  \bibinfo {pages} {239--242} (\bibinfo {year} {1999})}\BibitemShut {NoStop}%
\bibitem [{\citenamefont {Grangier}\ \emph {et~al.}(1998)\citenamefont
  {Grangier}, \citenamefont {Levenson},\ and\ \citenamefont
  {Poizat}}]{Grangier.98.N}%
  \BibitemOpen
  \bibfield  {author} {\bibinfo {author} {\bibfnamefont {P.}~\bibnamefont
  {Grangier}}, \bibinfo {author} {\bibfnamefont {J.~A.}\ \bibnamefont
  {Levenson}}, \ and\ \bibinfo {author} {\bibfnamefont {J-P.}\ \bibnamefont
  {Poizat}},\ }\bibfield  {title} {\enquote {\bibinfo {title} {Quantum
  non-demolition measurements in optics},}\ }\href
  {https://doi.org/10.1038/25059} {\bibfield  {journal} {\bibinfo  {journal}
  {Nature}\ }\textbf {\bibinfo {volume} {396}},\ \bibinfo {pages} {537--542}
  (\bibinfo {year} {1998})}\BibitemShut {NoStop}%
\bibitem [{\citenamefont {Kok}\ \emph {et~al.}(2002)\citenamefont {Kok},
  \citenamefont {Lee},\ and\ \citenamefont {Dowling}}]{PhysRevA.66.063814}%
  \BibitemOpen
  \bibfield  {author} {\bibinfo {author} {\bibfnamefont {P.}~\bibnamefont
  {Kok}}, \bibinfo {author} {\bibfnamefont {H.}~\bibnamefont {Lee}}, \ and\
  \bibinfo {author} {\bibfnamefont {J.~P.}\ \bibnamefont {Dowling}},\
  }\bibfield  {title} {\enquote {\bibinfo {title} {Single-photon
  quantum-nondemolition detectors constructed with linear optics and projective
  measurements},}\ }\href {\doibase 10.1103/PhysRevA.66.063814} {\bibfield
  {journal} {\bibinfo  {journal} {Phys. Rev. A}\ }\textbf {\bibinfo {volume}
  {66}},\ \bibinfo {pages} {063814} (\bibinfo {year} {2002})}\BibitemShut
  {NoStop}%
\bibitem [{\citenamefont {Guerlin}\ \emph {et~al.}(2007)\citenamefont
  {Guerlin}, \citenamefont {Bernu}, \citenamefont {Del\'eglise}, \citenamefont
  {Sayrin}, \citenamefont {Gleyzes}, \citenamefont {Kuhr}, \citenamefont
  {Brune}, \citenamefont {Raimond},\ and\ \citenamefont
  {Haroche}}]{Guerlin.07.N}%
  \BibitemOpen
  \bibfield  {author} {\bibinfo {author} {\bibfnamefont {C.}~\bibnamefont
  {Guerlin}}, \bibinfo {author} {\bibfnamefont {J.}~\bibnamefont {Bernu}},
  \bibinfo {author} {\bibfnamefont {S.}~\bibnamefont {Del\'eglise}}, \bibinfo
  {author} {\bibfnamefont {C.}~\bibnamefont {Sayrin}}, \bibinfo {author}
  {\bibfnamefont {S.}~\bibnamefont {Gleyzes}}, \bibinfo {author} {\bibfnamefont
  {S.}~\bibnamefont {Kuhr}}, \bibinfo {author} {\bibfnamefont {M.}~\bibnamefont
  {Brune}}, \bibinfo {author} {\bibfnamefont {J-M.}\ \bibnamefont {Raimond}}, \
  and\ \bibinfo {author} {\bibfnamefont {S.}~\bibnamefont {Haroche}},\
  }\bibfield  {title} {\enquote {\bibinfo {title} {Progressive field-state
  collapse and quantum non-demolition photon counting},}\ }\href
  {https://doi.org/10.1038/nature06057} {\bibfield  {journal} {\bibinfo
  {journal} {Nature}\ }\textbf {\bibinfo {volume} {448}},\ \bibinfo {pages}
  {889--893} (\bibinfo {year} {2007})}\BibitemShut {NoStop}%
\bibitem [{\citenamefont {Schuster}\ \emph {et~al.}(2007)\citenamefont
  {Schuster}, \citenamefont {Houck}, \citenamefont {Schreier}, \citenamefont
  {Wallraff}, \citenamefont {Gambetta}, \citenamefont {Blais}, \citenamefont
  {Frunzio}, \citenamefont {Majer}, \citenamefont {Johnson}, \citenamefont
  {Devoret}, \citenamefont {Girvin},\ and\ \citenamefont
  {Schoelkopf}}]{Schuster.07.N}%
  \BibitemOpen
  \bibfield  {author} {\bibinfo {author} {\bibfnamefont {D.~I.}\ \bibnamefont
  {Schuster}}, \bibinfo {author} {\bibfnamefont {A.~A.}\ \bibnamefont {Houck}},
  \bibinfo {author} {\bibfnamefont {J.~A.}\ \bibnamefont {Schreier}}, \bibinfo
  {author} {\bibfnamefont {A.}~\bibnamefont {Wallraff}}, \bibinfo {author}
  {\bibfnamefont {J.~M.}\ \bibnamefont {Gambetta}}, \bibinfo {author}
  {\bibfnamefont {A.}~\bibnamefont {Blais}}, \bibinfo {author} {\bibfnamefont
  {L.}~\bibnamefont {Frunzio}}, \bibinfo {author} {\bibfnamefont
  {J.}~\bibnamefont {Majer}}, \bibinfo {author} {\bibfnamefont
  {B.}~\bibnamefont {Johnson}}, \bibinfo {author} {\bibfnamefont {M.~H.}\
  \bibnamefont {Devoret}}, \bibinfo {author} {\bibfnamefont {S.~M.}\
  \bibnamefont {Girvin}}, \ and\ \bibinfo {author} {\bibfnamefont {R.~J.}\
  \bibnamefont {Schoelkopf}},\ }\bibfield  {title} {\enquote {\bibinfo {title}
  {Resolving photon number states in a superconducting circuit},}\ }\href
  {https://doi.org/10.1038/nature05461} {\bibfield  {journal} {\bibinfo
  {journal} {Nature}\ }\textbf {\bibinfo {volume} {445}},\ \bibinfo {pages}
  {515--518} (\bibinfo {year} {2007})}\BibitemShut {NoStop}%
\bibitem [{\citenamefont {Haroche}\ \emph {et~al.}(2009)\citenamefont
  {Haroche}, \citenamefont {Dotsenko}, \citenamefont {Del{\'{e}}glise},
  \citenamefont {Sayrin}, \citenamefont {Zhou}, \citenamefont {Gleyzes},
  \citenamefont {Guerlin}, \citenamefont {Kuhr}, \citenamefont {Brune},\ and\
  \citenamefont {Raimond}}]{Haroche.PS.09}%
  \BibitemOpen
  \bibfield  {author} {\bibinfo {author} {\bibfnamefont {S.}~\bibnamefont
  {Haroche}}, \bibinfo {author} {\bibfnamefont {I.}~\bibnamefont {Dotsenko}},
  \bibinfo {author} {\bibfnamefont {S.}~\bibnamefont {Del{\'{e}}glise}},
  \bibinfo {author} {\bibfnamefont {C.}~\bibnamefont {Sayrin}}, \bibinfo
  {author} {\bibfnamefont {X.}~\bibnamefont {Zhou}}, \bibinfo {author}
  {\bibfnamefont {S.}~\bibnamefont {Gleyzes}}, \bibinfo {author} {\bibfnamefont
  {C.}~\bibnamefont {Guerlin}}, \bibinfo {author} {\bibfnamefont
  {S.}~\bibnamefont {Kuhr}}, \bibinfo {author} {\bibfnamefont {M.}~\bibnamefont
  {Brune}}, \ and\ \bibinfo {author} {\bibfnamefont {J-M.}\ \bibnamefont
  {Raimond}},\ }\bibfield  {title} {\enquote {\bibinfo {title} {Manipulating
  and probing microwave fields in a cavity by quantum non-demolition photon
  counting},}\ }\href {\doibase 10.1088/0031-8949/2009/t137/014014} {\bibfield
  {journal} {\bibinfo  {journal} {Phys. Scr.}\ }\textbf {\bibinfo {volume}
  {T137}},\ \bibinfo {pages} {014014} (\bibinfo {year} {2009})}\BibitemShut
  {NoStop}%
\bibitem [{\citenamefont {Johnson}\ \emph {et~al.}(2010)\citenamefont
  {Johnson}, \citenamefont {Reed}, \citenamefont {Houck}, \citenamefont
  {Schuster}, \citenamefont {Bishop}, \citenamefont {Ginossar}, \citenamefont
  {Gambetta}, \citenamefont {DiCarlo}, \citenamefont {Frunzio}, \citenamefont
  {Girvin},\ and\ \citenamefont {Schoelkopf}}]{Johnson.10.NP}%
  \BibitemOpen
  \bibfield  {author} {\bibinfo {author} {\bibfnamefont {B.~R.}\ \bibnamefont
  {Johnson}}, \bibinfo {author} {\bibfnamefont {M.~D.}\ \bibnamefont {Reed}},
  \bibinfo {author} {\bibfnamefont {A.~A.}\ \bibnamefont {Houck}}, \bibinfo
  {author} {\bibfnamefont {D.~I.}\ \bibnamefont {Schuster}}, \bibinfo {author}
  {\bibfnamefont {Lev~S.}\ \bibnamefont {Bishop}}, \bibinfo {author}
  {\bibfnamefont {E.}~\bibnamefont {Ginossar}}, \bibinfo {author}
  {\bibfnamefont {J.~M.}\ \bibnamefont {Gambetta}}, \bibinfo {author}
  {\bibfnamefont {L.}~\bibnamefont {DiCarlo}}, \bibinfo {author} {\bibfnamefont
  {L.}~\bibnamefont {Frunzio}}, \bibinfo {author} {\bibfnamefont {S.~M.}\
  \bibnamefont {Girvin}}, \ and\ \bibinfo {author} {\bibfnamefont {R.~J.}\
  \bibnamefont {Schoelkopf}},\ }\bibfield  {title} {\enquote {\bibinfo {title}
  {Quantum non-demolition detection of single microwave photons in a
  circuit},}\ }\href {https://doi.org/10.1038/nphys1710} {\bibfield  {journal}
  {\bibinfo  {journal} {Nat. Phys.}\ }\textbf {\bibinfo {volume} {6}},\
  \bibinfo {pages} {663--667} (\bibinfo {year} {2010})}\BibitemShut {NoStop}%
\bibitem [{\citenamefont {Ludwig}\ \emph {et~al.}(2012)\citenamefont {Ludwig},
  \citenamefont {Safavi-Naeini}, \citenamefont {Painter},\ and\ \citenamefont
  {Marquardt}}]{Ludwig.12.PRL}%
  \BibitemOpen
  \bibfield  {author} {\bibinfo {author} {\bibfnamefont {M.}~\bibnamefont
  {Ludwig}}, \bibinfo {author} {\bibfnamefont {A.~H.}\ \bibnamefont
  {Safavi-Naeini}}, \bibinfo {author} {\bibfnamefont {O.}~\bibnamefont
  {Painter}}, \ and\ \bibinfo {author} {\bibfnamefont {F.}~\bibnamefont
  {Marquardt}},\ }\bibfield  {title} {\enquote {\bibinfo {title} {Enhanced
  quantum nonlinearities in a two-mode optomechanical system},}\ }\href
  {\doibase 10.1103/PhysRevLett.109.063601} {\bibfield  {journal} {\bibinfo
  {journal} {Phys. Rev. Lett.}\ }\textbf {\bibinfo {volume} {109}},\ \bibinfo
  {pages} {063601} (\bibinfo {year} {2012})}\BibitemShut {NoStop}%
\bibitem [{\citenamefont {Reiserer}\ \emph {et~al.}(2013)\citenamefont
  {Reiserer}, \citenamefont {Ritter},\ and\ \citenamefont
  {Rempe}}]{Reiserer.S.13}%
  \BibitemOpen
  \bibfield  {author} {\bibinfo {author} {\bibfnamefont {A.}~\bibnamefont
  {Reiserer}}, \bibinfo {author} {\bibfnamefont {S.}~\bibnamefont {Ritter}}, \
  and\ \bibinfo {author} {\bibfnamefont {G.}~\bibnamefont {Rempe}},\ }\bibfield
   {title} {\enquote {\bibinfo {title} {Nondestructive detection of an optical
  photon},}\ }\href {\doibase 10.1126/science.1246164} {\bibfield  {journal}
  {\bibinfo  {journal} {Science}\ }\textbf {\bibinfo {volume} {342}},\ \bibinfo
  {pages} {1349--1351} (\bibinfo {year} {2013})}\BibitemShut {NoStop}%
\bibitem [{\citenamefont {Peaudecerf}\ \emph {et~al.}(2014)\citenamefont
  {Peaudecerf}, \citenamefont {Rybarczyk}, \citenamefont {Gerlich},
  \citenamefont {Gleyzes}, \citenamefont {Raimond}, \citenamefont {Haroche},
  \citenamefont {Dotsenko},\ and\ \citenamefont {Brune}}]{Peaudecerf.14.PRL}%
  \BibitemOpen
  \bibfield  {author} {\bibinfo {author} {\bibfnamefont {B.}~\bibnamefont
  {Peaudecerf}}, \bibinfo {author} {\bibfnamefont {T.}~\bibnamefont
  {Rybarczyk}}, \bibinfo {author} {\bibfnamefont {S.}~\bibnamefont {Gerlich}},
  \bibinfo {author} {\bibfnamefont {S.}~\bibnamefont {Gleyzes}}, \bibinfo
  {author} {\bibfnamefont {J.~M.}\ \bibnamefont {Raimond}}, \bibinfo {author}
  {\bibfnamefont {S.}~\bibnamefont {Haroche}}, \bibinfo {author} {\bibfnamefont
  {I.}~\bibnamefont {Dotsenko}}, \ and\ \bibinfo {author} {\bibfnamefont
  {M.}~\bibnamefont {Brune}},\ }\bibfield  {title} {\enquote {\bibinfo {title}
  {Adaptive quantum nondemolition measurement of a photon number},}\ }\href
  {\doibase 10.1103/PhysRevLett.112.080401} {\bibfield  {journal} {\bibinfo
  {journal} {Phys. Rev. Lett.}\ }\textbf {\bibinfo {volume} {112}},\ \bibinfo
  {pages} {080401} (\bibinfo {year} {2014})}\BibitemShut {NoStop}%
\bibitem [{\citenamefont {Besse}\ \emph {et~al.}(2018)\citenamefont {Besse},
  \citenamefont {Gasparinetti}, \citenamefont {Collodo}, \citenamefont
  {Walter}, \citenamefont {Kurpiers}, \citenamefont {Pechal}, \citenamefont
  {Eichler},\ and\ \citenamefont {Wallraff}}]{Besse.18.PRX}%
  \BibitemOpen
  \bibfield  {author} {\bibinfo {author} {\bibfnamefont {J.}~\bibnamefont
  {Besse}}, \bibinfo {author} {\bibfnamefont {S.}~\bibnamefont {Gasparinetti}},
  \bibinfo {author} {\bibfnamefont {M.~C.}\ \bibnamefont {Collodo}}, \bibinfo
  {author} {\bibfnamefont {T.}~\bibnamefont {Walter}}, \bibinfo {author}
  {\bibfnamefont {P.}~\bibnamefont {Kurpiers}}, \bibinfo {author}
  {\bibfnamefont {M.}~\bibnamefont {Pechal}}, \bibinfo {author} {\bibfnamefont
  {C.}~\bibnamefont {Eichler}}, \ and\ \bibinfo {author} {\bibfnamefont
  {A.}~\bibnamefont {Wallraff}},\ }\bibfield  {title} {\enquote {\bibinfo
  {title} {Single-shot quantum nondemolition detection of individual itinerant
  microwave photons},}\ }\href {\doibase 10.1103/PhysRevX.8.021003} {\bibfield
  {journal} {\bibinfo  {journal} {Phys. Rev. X}\ }\textbf {\bibinfo {volume}
  {8}},\ \bibinfo {pages} {021003} (\bibinfo {year} {2018})}\BibitemShut
  {NoStop}%
\bibitem [{\citenamefont {Kono}\ \emph {et~al.}(2018)\citenamefont {Kono},
  \citenamefont {Koshino}, \citenamefont {Tabuchi}, \citenamefont {Noguchi},\
  and\ \citenamefont {Nakamura}}]{Kono.18.NP}%
  \BibitemOpen
  \bibfield  {author} {\bibinfo {author} {\bibfnamefont {S.}~\bibnamefont
  {Kono}}, \bibinfo {author} {\bibfnamefont {K.}~\bibnamefont {Koshino}},
  \bibinfo {author} {\bibfnamefont {Y.}~\bibnamefont {Tabuchi}}, \bibinfo
  {author} {\bibfnamefont {A.}~\bibnamefont {Noguchi}}, \ and\ \bibinfo
  {author} {\bibfnamefont {Y.}~\bibnamefont {Nakamura}},\ }\bibfield  {title}
  {\enquote {\bibinfo {title} {Quantum non-demolition detection of an itinerant
  microwave photon},}\ }\href {https://doi.org/10.1038/s41567-018-0066-3}
  {\bibfield  {journal} {\bibinfo  {journal} {Nat. Phys.}\ }\textbf {\bibinfo
  {volume} {14}},\ \bibinfo {pages} {546--549} (\bibinfo {year}
  {2018})}\BibitemShut {NoStop}%
\bibitem [{\citenamefont {Grimsmo}\ \emph {et~al.}(2020)\citenamefont
  {Grimsmo}, \citenamefont {Royer}, \citenamefont {Kreikebaum}, \citenamefont
  {Ye}, \citenamefont {O?Brien}, \citenamefont {Siddiqi},\ and\ \citenamefont
  {Blais}}]{Grimsmo.20.A}%
  \BibitemOpen
  \bibfield  {author} {\bibinfo {author} {\bibfnamefont {A.~L.}\ \bibnamefont
  {Grimsmo}}, \bibinfo {author} {\bibfnamefont {B.}~\bibnamefont {Royer}},
  \bibinfo {author} {\bibfnamefont {J.~M.}\ \bibnamefont {Kreikebaum}},
  \bibinfo {author} {\bibfnamefont {Y.}~\bibnamefont {Ye}}, \bibinfo {author}
  {\bibfnamefont {K.}~\bibnamefont {O?Brien}}, \bibinfo {author}
  {\bibfnamefont {I.}~\bibnamefont {Siddiqi}}, \ and\ \bibinfo {author}
  {\bibfnamefont {A.}~\bibnamefont {Blais}},\ }\bibfield  {title} {\enquote
  {\bibinfo {title} {Quantum metamaterial for nondestructive microwave photon
  counting},}\ }\href {https://arxiv.org/abs/2005.06483} {\  (\bibinfo {year}
  {2020})},\ \bibinfo {note} {arXiv:2005.06483}\BibitemShut {NoStop}%
\bibitem [{\citenamefont {Imoto}\ \emph {et~al.}(1985)\citenamefont {Imoto},
  \citenamefont {Haus},\ and\ \citenamefont {Yamamoto}}]{Imoto.85.PRA}%
  \BibitemOpen
  \bibfield  {author} {\bibinfo {author} {\bibfnamefont {N.}~\bibnamefont
  {Imoto}}, \bibinfo {author} {\bibfnamefont {H.~A.}\ \bibnamefont {Haus}}, \
  and\ \bibinfo {author} {\bibfnamefont {Y.}~\bibnamefont {Yamamoto}},\
  }\bibfield  {title} {\enquote {\bibinfo {title} {Quantum nondemolition
  measurement of the photon number via the optical kerr effect},}\ }\href
  {\doibase 10.1103/PhysRevA.32.2287} {\bibfield  {journal} {\bibinfo
  {journal} {Phys. Rev. A}\ }\textbf {\bibinfo {volume} {32}},\ \bibinfo
  {pages} {2287--2292} (\bibinfo {year} {1985})}\BibitemShut {NoStop}%
\bibitem [{\citenamefont {Munro}\ \emph {et~al.}(2005)\citenamefont {Munro},
  \citenamefont {Nemoto}, \citenamefont {Beausoleil},\ and\ \citenamefont
  {Spiller}}]{Munro.05.PRA}%
  \BibitemOpen
  \bibfield  {author} {\bibinfo {author} {\bibfnamefont {W.~J.}\ \bibnamefont
  {Munro}}, \bibinfo {author} {\bibfnamefont {Kae}\ \bibnamefont {Nemoto}},
  \bibinfo {author} {\bibfnamefont {R.~G.}\ \bibnamefont {Beausoleil}}, \ and\
  \bibinfo {author} {\bibfnamefont {T.~P.}\ \bibnamefont {Spiller}},\
  }\bibfield  {title} {\enquote {\bibinfo {title} {High-efficiency
  quantum-nondemolition single-photon-number-resolving detector},}\ }\href
  {\doibase 10.1103/PhysRevA.71.033819} {\bibfield  {journal} {\bibinfo
  {journal} {Phys. Rev. A}\ }\textbf {\bibinfo {volume} {71}},\ \bibinfo
  {pages} {033819} (\bibinfo {year} {2005})}\BibitemShut {NoStop}%
\bibitem [{\citenamefont {Malz}\ and\ \citenamefont
  {Cirac}(2020)}]{Malz.20.PRR}%
  \BibitemOpen
  \bibfield  {author} {\bibinfo {author} {\bibfnamefont {D.}~\bibnamefont
  {Malz}}\ and\ \bibinfo {author} {\bibfnamefont {J.~I.}\ \bibnamefont
  {Cirac}},\ }\bibfield  {title} {\enquote {\bibinfo {title} {Nondestructive
  photon counting in waveguide qed},}\ }\href {\doibase
  10.1103/PhysRevResearch.2.033091} {\bibfield  {journal} {\bibinfo  {journal}
  {Phys. Rev. Research}\ }\textbf {\bibinfo {volume} {2}},\ \bibinfo {pages}
  {033091} (\bibinfo {year} {2020})}\BibitemShut {NoStop}%
\bibitem [{\citenamefont {Blais}\ \emph {et~al.}(2004)\citenamefont {Blais},
  \citenamefont {Huang}, \citenamefont {Wallraff}, \citenamefont {Girvin},\
  and\ \citenamefont {Schoelkopf}}]{Blais.04.PRA}%
  \BibitemOpen
  \bibfield  {author} {\bibinfo {author} {\bibfnamefont {A.}~\bibnamefont
  {Blais}}, \bibinfo {author} {\bibfnamefont {R.}~\bibnamefont {Huang}},
  \bibinfo {author} {\bibfnamefont {A.}~\bibnamefont {Wallraff}}, \bibinfo
  {author} {\bibfnamefont {S.~M.}\ \bibnamefont {Girvin}}, \ and\ \bibinfo
  {author} {\bibfnamefont {R.~J.}\ \bibnamefont {Schoelkopf}},\ }\bibfield
  {title} {\enquote {\bibinfo {title} {Cavity quantum electrodynamics for
  superconducting electrical circuits: An architecture for quantum
  computation},}\ }\href {\doibase 10.1103/PhysRevA.69.062320} {\bibfield
  {journal} {\bibinfo  {journal} {Phys. Rev. A}\ }\textbf {\bibinfo {volume}
  {69}},\ \bibinfo {pages} {062320} (\bibinfo {year} {2004})}\BibitemShut
  {NoStop}%
\bibitem [{\citenamefont {Boissonneault}\ \emph {et~al.}(2008)\citenamefont
  {Boissonneault}, \citenamefont {Gambetta},\ and\ \citenamefont
  {Blais}}]{PhysRevA.77.060305}%
  \BibitemOpen
  \bibfield  {author} {\bibinfo {author} {\bibfnamefont {M.}~\bibnamefont
  {Boissonneault}}, \bibinfo {author} {\bibfnamefont {J.~M.}\ \bibnamefont
  {Gambetta}}, \ and\ \bibinfo {author} {\bibfnamefont {A.}~\bibnamefont
  {Blais}},\ }\bibfield  {title} {\enquote {\bibinfo {title} {Nonlinear
  dispersive regime of cavity qed: The dressed dephasing model},}\ }\href
  {\doibase 10.1103/PhysRevA.77.060305} {\bibfield  {journal} {\bibinfo
  {journal} {Phys. Rev. A}\ }\textbf {\bibinfo {volume} {77}},\ \bibinfo
  {pages} {060305} (\bibinfo {year} {2008})}\BibitemShut {NoStop}%
\bibitem [{\citenamefont {Slichter}\ \emph {et~al.}(2012)\citenamefont
  {Slichter}, \citenamefont {Vijay}, \citenamefont {Weber}, \citenamefont
  {Boutin}, \citenamefont {Boissonneault}, \citenamefont {Gambetta},
  \citenamefont {Blais},\ and\ \citenamefont {Siddiqi}}]{Slichter.12.PRL}%
  \BibitemOpen
  \bibfield  {author} {\bibinfo {author} {\bibfnamefont {D.~H.}\ \bibnamefont
  {Slichter}}, \bibinfo {author} {\bibfnamefont {R.}~\bibnamefont {Vijay}},
  \bibinfo {author} {\bibfnamefont {S.~J.}\ \bibnamefont {Weber}}, \bibinfo
  {author} {\bibfnamefont {S.}~\bibnamefont {Boutin}}, \bibinfo {author}
  {\bibfnamefont {M.}~\bibnamefont {Boissonneault}}, \bibinfo {author}
  {\bibfnamefont {J.~M.}\ \bibnamefont {Gambetta}}, \bibinfo {author}
  {\bibfnamefont {A.}~\bibnamefont {Blais}}, \ and\ \bibinfo {author}
  {\bibfnamefont {I.}~\bibnamefont {Siddiqi}},\ }\bibfield  {title} {\enquote
  {\bibinfo {title} {Measurement-induced qubit state mixing in circuit qed from
  up-converted dephasing noise},}\ }\href {\doibase
  10.1103/PhysRevLett.109.153601} {\bibfield  {journal} {\bibinfo  {journal}
  {Phys. Rev. Lett.}\ }\textbf {\bibinfo {volume} {109}},\ \bibinfo {pages}
  {153601} (\bibinfo {year} {2012})}\BibitemShut {NoStop}%
\bibitem [{\citenamefont {Sank}\ \emph {et~al.}(2016)\citenamefont {Sank},
  \citenamefont {Chen}, \citenamefont {Khezri}, \citenamefont {Kelly},
  \citenamefont {Barends}, \citenamefont {Campbell}, \citenamefont {Chen},
  \citenamefont {Chiaro}, \citenamefont {Dunsworth}, \citenamefont {Fowler},
  \citenamefont {Jeffrey}, \citenamefont {Lucero}, \citenamefont {Megrant},
  \citenamefont {Mutus}, \citenamefont {Neeley}, \citenamefont {Neill},
  \citenamefont {O'Malley}, \citenamefont {Quintana}, \citenamefont {Roushan},
  \citenamefont {Vainsencher}, \citenamefont {White}, \citenamefont {Wenner},
  \citenamefont {Korotkov},\ and\ \citenamefont {Martinis}}]{Sank.16.PRL}%
  \BibitemOpen
  \bibfield  {author} {\bibinfo {author} {\bibfnamefont {D.}~\bibnamefont
  {Sank}}, \bibinfo {author} {\bibfnamefont {Z.}~\bibnamefont {Chen}}, \bibinfo
  {author} {\bibfnamefont {M.}~\bibnamefont {Khezri}}, \bibinfo {author}
  {\bibfnamefont {J.}~\bibnamefont {Kelly}}, \bibinfo {author} {\bibfnamefont
  {R.}~\bibnamefont {Barends}}, \bibinfo {author} {\bibfnamefont
  {B.}~\bibnamefont {Campbell}}, \bibinfo {author} {\bibfnamefont
  {Y.}~\bibnamefont {Chen}}, \bibinfo {author} {\bibfnamefont {B.}~\bibnamefont
  {Chiaro}}, \bibinfo {author} {\bibfnamefont {A.}~\bibnamefont {Dunsworth}},
  \bibinfo {author} {\bibfnamefont {A.}~\bibnamefont {Fowler}}, \bibinfo
  {author} {\bibfnamefont {E.}~\bibnamefont {Jeffrey}}, \bibinfo {author}
  {\bibfnamefont {E.}~\bibnamefont {Lucero}}, \bibinfo {author} {\bibfnamefont
  {A.}~\bibnamefont {Megrant}}, \bibinfo {author} {\bibfnamefont
  {J.}~\bibnamefont {Mutus}}, \bibinfo {author} {\bibfnamefont
  {M.}~\bibnamefont {Neeley}}, \bibinfo {author} {\bibfnamefont
  {C.}~\bibnamefont {Neill}}, \bibinfo {author} {\bibfnamefont {P.~J.~J.}\
  \bibnamefont {O'Malley}}, \bibinfo {author} {\bibfnamefont {C.}~\bibnamefont
  {Quintana}}, \bibinfo {author} {\bibfnamefont {P.}~\bibnamefont {Roushan}},
  \bibinfo {author} {\bibfnamefont {A.}~\bibnamefont {Vainsencher}}, \bibinfo
  {author} {\bibfnamefont {T.}~\bibnamefont {White}}, \bibinfo {author}
  {\bibfnamefont {J.}~\bibnamefont {Wenner}}, \bibinfo {author} {\bibfnamefont
  {Alexander~N.}\ \bibnamefont {Korotkov}}, \ and\ \bibinfo {author}
  {\bibfnamefont {J.~M.}\ \bibnamefont {Martinis}},\ }\bibfield  {title}
  {\enquote {\bibinfo {title} {Measurement-induced state transitions in a
  superconducting qubit: Beyond the rotating wave approximation},}\ }\href
  {\doibase 10.1103/PhysRevLett.117.190503} {\bibfield  {journal} {\bibinfo
  {journal} {Phys. Rev. Lett.}\ }\textbf {\bibinfo {volume} {117}},\ \bibinfo
  {pages} {190503} (\bibinfo {year} {2016})}\BibitemShut {NoStop}%
\bibitem [{\citenamefont {Dassonneville}\ \emph {et~al.}(2020)\citenamefont
  {Dassonneville}, \citenamefont {Ramos}, \citenamefont {Milchakov},
  \citenamefont {Planat}, \citenamefont {Dumur}, \citenamefont {Foroughi},
  \citenamefont {Puertas}, \citenamefont {Leger}, \citenamefont {Bharadwaj},
  \citenamefont {Delaforce}, \citenamefont {Naud}, \citenamefont
  {Hasch-Guichard}, \citenamefont {Garc\'{\i}a-Ripoll}, \citenamefont {Roch},\
  and\ \citenamefont {Buisson}}]{Dassonneville.20.PRX}%
  \BibitemOpen
  \bibfield  {author} {\bibinfo {author} {\bibfnamefont {R.}~\bibnamefont
  {Dassonneville}}, \bibinfo {author} {\bibfnamefont {T.}~\bibnamefont
  {Ramos}}, \bibinfo {author} {\bibfnamefont {V.}~\bibnamefont {Milchakov}},
  \bibinfo {author} {\bibfnamefont {L.}~\bibnamefont {Planat}}, \bibinfo
  {author} {\bibfnamefont {\'E.}\ \bibnamefont {Dumur}}, \bibinfo {author}
  {\bibfnamefont {F.}~\bibnamefont {Foroughi}}, \bibinfo {author}
  {\bibfnamefont {J.}~\bibnamefont {Puertas}}, \bibinfo {author} {\bibfnamefont
  {S.}~\bibnamefont {Leger}}, \bibinfo {author} {\bibfnamefont
  {K.}~\bibnamefont {Bharadwaj}}, \bibinfo {author} {\bibfnamefont
  {J.}~\bibnamefont {Delaforce}}, \bibinfo {author} {\bibfnamefont
  {C.}~\bibnamefont {Naud}}, \bibinfo {author} {\bibfnamefont {W.}~\bibnamefont
  {Hasch-Guichard}}, \bibinfo {author} {\bibfnamefont {J.~J.}\ \bibnamefont
  {Garc\'{\i}a-Ripoll}}, \bibinfo {author} {\bibfnamefont {N.}~\bibnamefont
  {Roch}}, \ and\ \bibinfo {author} {\bibfnamefont {O.}~\bibnamefont
  {Buisson}},\ }\bibfield  {title} {\enquote {\bibinfo {title} {Fast
  high-fidelity quantum nondemolition qubit readout via a nonperturbative
  cross-kerr coupling},}\ }\href {\doibase 10.1103/PhysRevX.10.011045}
  {\bibfield  {journal} {\bibinfo  {journal} {Phys. Rev. X}\ }\textbf {\bibinfo
  {volume} {10}},\ \bibinfo {pages} {011045} (\bibinfo {year}
  {2020})}\BibitemShut {NoStop}%
\bibitem [{\citenamefont {Knobel}\ and\ \citenamefont
  {Cleland}(2003)}]{Knobel.03.N}%
  \BibitemOpen
  \bibfield  {author} {\bibinfo {author} {\bibfnamefont {R.~G.}\ \bibnamefont
  {Knobel}}\ and\ \bibinfo {author} {\bibfnamefont {A.~N.}\ \bibnamefont
  {Cleland}},\ }\bibfield  {title} {\enquote {\bibinfo {title} {Nanometre-scale
  displacement sensing using a single electron transistor},}\ }\href
  {https://doi.org/10.1038/nature01773} {\bibfield  {journal} {\bibinfo
  {journal} {Nature}\ }\textbf {\bibinfo {volume} {424}},\ \bibinfo {pages}
  {291--293} (\bibinfo {year} {2003})}\BibitemShut {NoStop}%
\bibitem [{\citenamefont {LaHaye}\ \emph {et~al.}(2004)\citenamefont {LaHaye},
  \citenamefont {Buu}, \citenamefont {Camarota},\ and\ \citenamefont
  {Schwab}}]{LaHaye.S.04}%
  \BibitemOpen
  \bibfield  {author} {\bibinfo {author} {\bibfnamefont {M.~D.}\ \bibnamefont
  {LaHaye}}, \bibinfo {author} {\bibfnamefont {O.}~\bibnamefont {Buu}},
  \bibinfo {author} {\bibfnamefont {B.}~\bibnamefont {Camarota}}, \ and\
  \bibinfo {author} {\bibfnamefont {K.~C.}\ \bibnamefont {Schwab}},\ }\bibfield
   {title} {\enquote {\bibinfo {title} {Approaching the quantum limit of a
  nanomechanical resonator},}\ }\href {\doibase 10.1126/science.1094419}
  {\bibfield  {journal} {\bibinfo  {journal} {Science}\ }\textbf {\bibinfo
  {volume} {304}},\ \bibinfo {pages} {74--77} (\bibinfo {year}
  {2004})}\BibitemShut {NoStop}%
\bibitem [{\citenamefont {Mozyrsky}\ \emph {et~al.}(2004)\citenamefont
  {Mozyrsky}, \citenamefont {Martin},\ and\ \citenamefont
  {Hastings}}]{Mozyrsky.04.PRL}%
  \BibitemOpen
  \bibfield  {author} {\bibinfo {author} {\bibfnamefont {D.}~\bibnamefont
  {Mozyrsky}}, \bibinfo {author} {\bibfnamefont {I.}~\bibnamefont {Martin}}, \
  and\ \bibinfo {author} {\bibfnamefont {M.~B.}\ \bibnamefont {Hastings}},\
  }\bibfield  {title} {\enquote {\bibinfo {title} {Quantum-limited sensitivity
  of single-electron-transistor-based displacement detectors},}\ }\href
  {\doibase 10.1103/PhysRevLett.92.018303} {\bibfield  {journal} {\bibinfo
  {journal} {Phys. Rev. Lett.}\ }\textbf {\bibinfo {volume} {92}},\ \bibinfo
  {pages} {018303} (\bibinfo {year} {2004})}\BibitemShut {NoStop}%
\bibitem [{\citenamefont {Blencowe}(2004)}]{Blencowe.04.PR}%
  \BibitemOpen
  \bibfield  {author} {\bibinfo {author} {\bibfnamefont {M.}~\bibnamefont
  {Blencowe}},\ }\bibfield  {title} {\enquote {\bibinfo {title} {Quantum
  electromechanical systems},}\ }\href {\doibase
  https://doi.org/10.1016/j.physrep.2003.12.005} {\bibfield  {journal}
  {\bibinfo  {journal} {Phys. Rep.}\ }\textbf {\bibinfo {volume} {395}},\
  \bibinfo {pages} {159 -- 222} (\bibinfo {year} {2004})}\BibitemShut {NoStop}%
\bibitem [{\citenamefont {Aspelmeyer}\ \emph {et~al.}(2014)\citenamefont
  {Aspelmeyer}, \citenamefont {Kippenberg},\ and\ \citenamefont
  {Marquardt}}]{Aspelmeyer.14.RMP}%
  \BibitemOpen
  \bibfield  {author} {\bibinfo {author} {\bibfnamefont {M.}~\bibnamefont
  {Aspelmeyer}}, \bibinfo {author} {\bibfnamefont {T.~J.}\ \bibnamefont
  {Kippenberg}}, \ and\ \bibinfo {author} {\bibfnamefont {F.}~\bibnamefont
  {Marquardt}},\ }\bibfield  {title} {\enquote {\bibinfo {title} {Cavity
  optomechanics},}\ }\href {\doibase 10.1103/RevModPhys.86.1391} {\bibfield
  {journal} {\bibinfo  {journal} {Rev. Mod. Phys.}\ }\textbf {\bibinfo {volume}
  {86}},\ \bibinfo {pages} {1391--1452} (\bibinfo {year} {2014})}\BibitemShut
  {NoStop}%
\bibitem [{\citenamefont {Steele}\ \emph {et~al.}(2009)\citenamefont {Steele},
  \citenamefont {H{\"u}ttel}, \citenamefont {Witkamp}, \citenamefont {Poot},
  \citenamefont {Meerwaldt}, \citenamefont {Kouwenhoven},\ and\ \citenamefont
  {van~der Zant}}]{Steele.S.09}%
  \BibitemOpen
  \bibfield  {author} {\bibinfo {author} {\bibfnamefont {G.~A.}\ \bibnamefont
  {Steele}}, \bibinfo {author} {\bibfnamefont {A.~K.}\ \bibnamefont
  {H{\"u}ttel}}, \bibinfo {author} {\bibfnamefont {B.}~\bibnamefont {Witkamp}},
  \bibinfo {author} {\bibfnamefont {M.}~\bibnamefont {Poot}}, \bibinfo {author}
  {\bibfnamefont {H.~B.}\ \bibnamefont {Meerwaldt}}, \bibinfo {author}
  {\bibfnamefont {L.~P.}\ \bibnamefont {Kouwenhoven}}, \ and\ \bibinfo {author}
  {\bibfnamefont {H.~S.~J.}\ \bibnamefont {van~der Zant}},\ }\bibfield  {title}
  {\enquote {\bibinfo {title} {Strong coupling between single-electron
  tunneling and nanomechanical motion},}\ }\href {\doibase
  10.1126/science.1176076} {\bibfield  {journal} {\bibinfo  {journal}
  {Science}\ }\textbf {\bibinfo {volume} {325}},\ \bibinfo {pages} {1103--1107}
  (\bibinfo {year} {2009})}\BibitemShut {NoStop}%
\bibitem [{\citenamefont {Wen}\ \emph {et~al.}(2020)\citenamefont {Wen},
  \citenamefont {Ares}, \citenamefont {Schupp}, \citenamefont {Pei},
  \citenamefont {Briggs},\ and\ \citenamefont {Laird}}]{Wen.20.NP}%
  \BibitemOpen
  \bibfield  {author} {\bibinfo {author} {\bibfnamefont {Y.}~\bibnamefont
  {Wen}}, \bibinfo {author} {\bibfnamefont {N.}~\bibnamefont {Ares}}, \bibinfo
  {author} {\bibfnamefont {F.~J.}\ \bibnamefont {Schupp}}, \bibinfo {author}
  {\bibfnamefont {T.}~\bibnamefont {Pei}}, \bibinfo {author} {\bibfnamefont
  {G.~A.~D.}\ \bibnamefont {Briggs}}, \ and\ \bibinfo {author} {\bibfnamefont
  {E.~A.}\ \bibnamefont {Laird}},\ }\bibfield  {title} {\enquote {\bibinfo
  {title} {A coherent nanomechanical oscillator driven by single-electron
  tunnelling},}\ }\href {https://doi.org/10.1038/s41567-019-0683-5} {\bibfield
  {journal} {\bibinfo  {journal} {Nat. Phys.}\ }\textbf {\bibinfo {volume}
  {16}},\ \bibinfo {pages} {75--82} (\bibinfo {year} {2020})}\BibitemShut
  {NoStop}%
\bibitem [{\citenamefont {Blien}\ \emph {et~al.}(2020)\citenamefont {Blien},
  \citenamefont {Steger}, \citenamefont {Hüttner}, \citenamefont {Graaf},\
  and\ \citenamefont {Hüttel}}]{Blien.20.NC}%
  \BibitemOpen
  \bibfield  {author} {\bibinfo {author} {\bibfnamefont {S.}~\bibnamefont
  {Blien}}, \bibinfo {author} {\bibfnamefont {P.}~\bibnamefont {Steger}},
  \bibinfo {author} {\bibfnamefont {N.}~\bibnamefont {Hüttner}}, \bibinfo
  {author} {\bibfnamefont {R.}~\bibnamefont {Graaf}}, \ and\ \bibinfo {author}
  {\bibfnamefont {A.~K.}\ \bibnamefont {Hüttel}},\ }\bibfield  {title}
  {\enquote {\bibinfo {title} {Quantum capacitance mediated carbon nanotube
  optomechanics},}\ }\href {https://doi.org/10.1038/s41467-020-15433-3}
  {\bibfield  {journal} {\bibinfo  {journal} {Nat. Commun.}\ }\textbf {\bibinfo
  {volume} {11}},\ \bibinfo {pages} {1636} (\bibinfo {year}
  {2020})}\BibitemShut {NoStop}%
\bibitem [{\citenamefont {Naik}\ \emph {et~al.}(2006)\citenamefont {Naik},
  \citenamefont {Buu}, \citenamefont {LaHaye}, \citenamefont {Armour},
  \citenamefont {Clerk}, \citenamefont {Blencowe},\ and\ \citenamefont
  {Schwab}}]{Naik.06.N}%
  \BibitemOpen
  \bibfield  {author} {\bibinfo {author} {\bibfnamefont {A.}~\bibnamefont
  {Naik}}, \bibinfo {author} {\bibfnamefont {O.}~\bibnamefont {Buu}}, \bibinfo
  {author} {\bibfnamefont {M.~D.}\ \bibnamefont {LaHaye}}, \bibinfo {author}
  {\bibfnamefont {A.~D.}\ \bibnamefont {Armour}}, \bibinfo {author}
  {\bibfnamefont {A.~A.}\ \bibnamefont {Clerk}}, \bibinfo {author}
  {\bibfnamefont {M.~P.}\ \bibnamefont {Blencowe}}, \ and\ \bibinfo {author}
  {\bibfnamefont {K.~C.}\ \bibnamefont {Schwab}},\ }\bibfield  {title}
  {\enquote {\bibinfo {title} {Cooling a nanomechanical resonator with quantum
  back-action},}\ }\href {https://doi.org/10.1038/nature05027} {\bibfield
  {journal} {\bibinfo  {journal} {Nature}\ }\textbf {\bibinfo {volume} {443}},\
  \bibinfo {pages} {193--196} (\bibinfo {year} {2006})}\BibitemShut {NoStop}%
\bibitem [{\citenamefont {Anetsberger}\ \emph {et~al.}(2009)\citenamefont
  {Anetsberger}, \citenamefont {Arcizet}, \citenamefont {Unterreithmeier},
  \citenamefont {Rivière}, \citenamefont {Schliesser}, \citenamefont {Weig},
  \citenamefont {Kotthaus},\ and\ \citenamefont
  {Kippenberg}}]{Anetsberger.09.NP}%
  \BibitemOpen
  \bibfield  {author} {\bibinfo {author} {\bibfnamefont {G.}~\bibnamefont
  {Anetsberger}}, \bibinfo {author} {\bibfnamefont {O.}~\bibnamefont
  {Arcizet}}, \bibinfo {author} {\bibfnamefont {Q.~P.}\ \bibnamefont
  {Unterreithmeier}}, \bibinfo {author} {\bibfnamefont {R.}~\bibnamefont
  {Rivière}}, \bibinfo {author} {\bibfnamefont {A.}~\bibnamefont
  {Schliesser}}, \bibinfo {author} {\bibfnamefont {E.~M.}\ \bibnamefont
  {Weig}}, \bibinfo {author} {\bibfnamefont {J.~P.}\ \bibnamefont {Kotthaus}},
  \ and\ \bibinfo {author} {\bibfnamefont {T.~J.}\ \bibnamefont {Kippenberg}},\
  }\bibfield  {title} {\enquote {\bibinfo {title} {Near-field cavity
  optomechanics with nanomechanical oscillators},}\ }\href
  {https://doi.org/10.1038/nphys1425} {\bibfield  {journal} {\bibinfo
  {journal} {Nat. Phys.}\ }\textbf {\bibinfo {volume} {5}},\ \bibinfo {pages}
  {909--914} (\bibinfo {year} {2009})}\BibitemShut {NoStop}%
\bibitem [{\citenamefont {Gr\"oblacher}\ \emph {et~al.}(2009)\citenamefont
  {Gr\"oblacher}, \citenamefont {Hammerer}, \citenamefont {Vanner},\ and\
  \citenamefont {Aspelmeyer}}]{Groblacher.09.N}%
  \BibitemOpen
  \bibfield  {author} {\bibinfo {author} {\bibfnamefont {S.}~\bibnamefont
  {Gr\"oblacher}}, \bibinfo {author} {\bibfnamefont {K.}~\bibnamefont
  {Hammerer}}, \bibinfo {author} {\bibfnamefont {M.~R.}\ \bibnamefont
  {Vanner}}, \ and\ \bibinfo {author} {\bibfnamefont {M.}~\bibnamefont
  {Aspelmeyer}},\ }\bibfield  {title} {\enquote {\bibinfo {title} {Observation
  of strong coupling between a micromechanical resonator and an optical cavity
  field},}\ }\href {https://doi.org/10.1038/nature08171} {\bibfield  {journal}
  {\bibinfo  {journal} {Nature}\ }\textbf {\bibinfo {volume} {460}},\ \bibinfo
  {pages} {724--727} (\bibinfo {year} {2009})}\BibitemShut {NoStop}%
\bibitem [{\citenamefont {Clerk}\ \emph {et~al.}(2010)\citenamefont {Clerk},
  \citenamefont {Devoret}, \citenamefont {Girvin}, \citenamefont {Marquardt},\
  and\ \citenamefont {Schoelkopf}}]{Clerk.10.RMP}%
  \BibitemOpen
  \bibfield  {author} {\bibinfo {author} {\bibfnamefont {A.~A.}\ \bibnamefont
  {Clerk}}, \bibinfo {author} {\bibfnamefont {M.~H.}\ \bibnamefont {Devoret}},
  \bibinfo {author} {\bibfnamefont {S.~M.}\ \bibnamefont {Girvin}}, \bibinfo
  {author} {\bibfnamefont {Florian}\ \bibnamefont {Marquardt}}, \ and\ \bibinfo
  {author} {\bibfnamefont {R.~J.}\ \bibnamefont {Schoelkopf}},\ }\bibfield
  {title} {\enquote {\bibinfo {title} {Introduction to quantum noise,
  measurement, and amplification},}\ }\href {\doibase
  10.1103/RevModPhys.82.1155} {\bibfield  {journal} {\bibinfo  {journal} {Rev.
  Mod. Phys.}\ }\textbf {\bibinfo {volume} {82}},\ \bibinfo {pages}
  {1155--1208} (\bibinfo {year} {2010})}\BibitemShut {NoStop}%
\bibitem [{\citenamefont {Stannigel}\ \emph {et~al.}(2010)\citenamefont
  {Stannigel}, \citenamefont {Rabl}, \citenamefont {S\o{}rensen}, \citenamefont
  {Zoller},\ and\ \citenamefont {Lukin}}]{Stannigel.10.PRL}%
  \BibitemOpen
  \bibfield  {author} {\bibinfo {author} {\bibfnamefont {K.}~\bibnamefont
  {Stannigel}}, \bibinfo {author} {\bibfnamefont {P.}~\bibnamefont {Rabl}},
  \bibinfo {author} {\bibfnamefont {A.~S.}\ \bibnamefont {S\o{}rensen}},
  \bibinfo {author} {\bibfnamefont {P.}~\bibnamefont {Zoller}}, \ and\ \bibinfo
  {author} {\bibfnamefont {M.~D.}\ \bibnamefont {Lukin}},\ }\bibfield  {title}
  {\enquote {\bibinfo {title} {Optomechanical transducers for long-distance
  quantum communication},}\ }\href {\doibase 10.1103/PhysRevLett.105.220501}
  {\bibfield  {journal} {\bibinfo  {journal} {Phys. Rev. Lett.}\ }\textbf
  {\bibinfo {volume} {105}},\ \bibinfo {pages} {220501} (\bibinfo {year}
  {2010})}\BibitemShut {NoStop}%
\bibitem [{\citenamefont {Teufel}\ \emph
  {et~al.}(2011{\natexlab{a}})\citenamefont {Teufel}, \citenamefont {Donner},
  \citenamefont {Li}, \citenamefont {Harlow}, \citenamefont {Allman},
  \citenamefont {Cicak}, \citenamefont {Sirois}, \citenamefont {Whittaker},
  \citenamefont {Lehnert},\ and\ \citenamefont {Simmonds}}]{Teufel.11.N}%
  \BibitemOpen
  \bibfield  {author} {\bibinfo {author} {\bibfnamefont {J.~D.}\ \bibnamefont
  {Teufel}}, \bibinfo {author} {\bibfnamefont {T.}~\bibnamefont {Donner}},
  \bibinfo {author} {\bibfnamefont {Dale}\ \bibnamefont {Li}}, \bibinfo
  {author} {\bibfnamefont {J.~W.}\ \bibnamefont {Harlow}}, \bibinfo {author}
  {\bibfnamefont {M.~S.}\ \bibnamefont {Allman}}, \bibinfo {author}
  {\bibfnamefont {K.}~\bibnamefont {Cicak}}, \bibinfo {author} {\bibfnamefont
  {A.~J.}\ \bibnamefont {Sirois}}, \bibinfo {author} {\bibfnamefont {J.~D.}\
  \bibnamefont {Whittaker}}, \bibinfo {author} {\bibfnamefont {K.~W.}\
  \bibnamefont {Lehnert}}, \ and\ \bibinfo {author} {\bibfnamefont {R.~W.}\
  \bibnamefont {Simmonds}},\ }\bibfield  {title} {\enquote {\bibinfo {title}
  {Sideband cooling of micromechanical motion to the quantum ground state},}\
  }\href {https://doi.org/10.1038/nature10261} {\bibfield  {journal} {\bibinfo
  {journal} {Nature}\ }\textbf {\bibinfo {volume} {475}},\ \bibinfo {pages}
  {359--363} (\bibinfo {year} {2011}{\natexlab{a}})}\BibitemShut {NoStop}%
\bibitem [{\citenamefont {Teufel}\ \emph
  {et~al.}(2011{\natexlab{b}})\citenamefont {Teufel}, \citenamefont {Li},
  \citenamefont {Allman}, \citenamefont {Cicak}, \citenamefont {Sirois},
  \citenamefont {Whittaker},\ and\ \citenamefont {Simmonds}}]{Teufel.11.Na}%
  \BibitemOpen
  \bibfield  {author} {\bibinfo {author} {\bibfnamefont {J.~D.}\ \bibnamefont
  {Teufel}}, \bibinfo {author} {\bibfnamefont {Dale}\ \bibnamefont {Li}},
  \bibinfo {author} {\bibfnamefont {M.~S.}\ \bibnamefont {Allman}}, \bibinfo
  {author} {\bibfnamefont {K.}~\bibnamefont {Cicak}}, \bibinfo {author}
  {\bibfnamefont {A.~J.}\ \bibnamefont {Sirois}}, \bibinfo {author}
  {\bibfnamefont {J.~D.}\ \bibnamefont {Whittaker}}, \ and\ \bibinfo {author}
  {\bibfnamefont {R.~W.}\ \bibnamefont {Simmonds}},\ }\bibfield  {title}
  {\enquote {\bibinfo {title} {Circuit cavity electromechanics in the
  strong-coupling regime},}\ }\href {https://doi.org/10.1038/nature09898}
  {\bibfield  {journal} {\bibinfo  {journal} {Nature}\ }\textbf {\bibinfo
  {volume} {471}},\ \bibinfo {pages} {204--208} (\bibinfo {year}
  {2011}{\natexlab{b}})}\BibitemShut {NoStop}%
\bibitem [{\citenamefont {Verhagen}\ \emph {et~al.}(2012)\citenamefont
  {Verhagen}, \citenamefont {Del\'eglise}, \citenamefont {Weis}, \citenamefont
  {Schliesser},\ and\ \citenamefont {Kippenberg}}]{Verhagen.12.N}%
  \BibitemOpen
  \bibfield  {author} {\bibinfo {author} {\bibfnamefont {E.}~\bibnamefont
  {Verhagen}}, \bibinfo {author} {\bibfnamefont {S.}~\bibnamefont
  {Del\'eglise}}, \bibinfo {author} {\bibfnamefont {S.}~\bibnamefont {Weis}},
  \bibinfo {author} {\bibfnamefont {A.}~\bibnamefont {Schliesser}}, \ and\
  \bibinfo {author} {\bibfnamefont {T.~J.}\ \bibnamefont {Kippenberg}},\
  }\bibfield  {title} {\enquote {\bibinfo {title} {Quantum-coherent coupling of
  a mechanical oscillator to an optical cavity mode},}\ }\href
  {https://doi.org/10.1038/nature10787} {\bibfield  {journal} {\bibinfo
  {journal} {Nature}\ }\textbf {\bibinfo {volume} {482}},\ \bibinfo {pages}
  {63--67} (\bibinfo {year} {2012})}\BibitemShut {NoStop}%
\bibitem [{\citenamefont {Xiang}\ \emph {et~al.}(2013)\citenamefont {Xiang},
  \citenamefont {Ashhab}, \citenamefont {You},\ and\ \citenamefont
  {Nori}}]{Xiang.13.RMP}%
  \BibitemOpen
  \bibfield  {author} {\bibinfo {author} {\bibfnamefont {Z.}~\bibnamefont
  {Xiang}}, \bibinfo {author} {\bibfnamefont {S.}~\bibnamefont {Ashhab}},
  \bibinfo {author} {\bibfnamefont {J.~Q.}\ \bibnamefont {You}}, \ and\
  \bibinfo {author} {\bibfnamefont {F.}~\bibnamefont {Nori}},\ }\bibfield
  {title} {\enquote {\bibinfo {title} {Hybrid quantum circuits: Superconducting
  circuits interacting with other quantum systems},}\ }\href {\doibase
  10.1103/RevModPhys.85.623} {\bibfield  {journal} {\bibinfo  {journal} {Rev.
  Mod. Phys.}\ }\textbf {\bibinfo {volume} {85}},\ \bibinfo {pages} {623--653}
  (\bibinfo {year} {2013})}\BibitemShut {NoStop}%
\bibitem [{\citenamefont {Lassagne}\ \emph {et~al.}(2009)\citenamefont
  {Lassagne}, \citenamefont {Tarakanov}, \citenamefont {Kinaret}, \citenamefont
  {Garcia-Sanchez},\ and\ \citenamefont {Bachtold}}]{Lassagne.S.09}%
  \BibitemOpen
  \bibfield  {author} {\bibinfo {author} {\bibfnamefont {B.}~\bibnamefont
  {Lassagne}}, \bibinfo {author} {\bibfnamefont {Y.}~\bibnamefont {Tarakanov}},
  \bibinfo {author} {\bibfnamefont {J.}~\bibnamefont {Kinaret}}, \bibinfo
  {author} {\bibfnamefont {D.}~\bibnamefont {Garcia-Sanchez}}, \ and\ \bibinfo
  {author} {\bibfnamefont {A.}~\bibnamefont {Bachtold}},\ }\bibfield  {title}
  {\enquote {\bibinfo {title} {Coupling mechanics to charge transport in carbon
  nanotube mechanical resonators},}\ }\href {\doibase 10.1126/science.1174290}
  {\bibfield  {journal} {\bibinfo  {journal} {Science}\ }\textbf {\bibinfo
  {volume} {325}},\ \bibinfo {pages} {1107--1110} (\bibinfo {year}
  {2009})}\BibitemShut {NoStop}%
\bibitem [{\citenamefont {Lee}\ \emph {et~al.}(2010)\citenamefont {Lee},
  \citenamefont {McRae}, \citenamefont {Harris}, \citenamefont {Knittel},\ and\
  \citenamefont {Bowen}}]{Lee.10.PRL}%
  \BibitemOpen
  \bibfield  {author} {\bibinfo {author} {\bibfnamefont {K.~H.}\ \bibnamefont
  {Lee}}, \bibinfo {author} {\bibfnamefont {T.~G.}\ \bibnamefont {McRae}},
  \bibinfo {author} {\bibfnamefont {G.~I.}\ \bibnamefont {Harris}}, \bibinfo
  {author} {\bibfnamefont {J.}~\bibnamefont {Knittel}}, \ and\ \bibinfo
  {author} {\bibfnamefont {W.~P.}\ \bibnamefont {Bowen}},\ }\bibfield  {title}
  {\enquote {\bibinfo {title} {Cooling and control of a cavity
  optoelectromechanical system},}\ }\href {\doibase
  10.1103/PhysRevLett.104.123604} {\bibfield  {journal} {\bibinfo  {journal}
  {Phys. Rev. Lett.}\ }\textbf {\bibinfo {volume} {104}},\ \bibinfo {pages}
  {123604} (\bibinfo {year} {2010})}\BibitemShut {NoStop}%
\bibitem [{\citenamefont {Winger}\ \emph {et~al.}(2011)\citenamefont {Winger},
  \citenamefont {Blasius}, \citenamefont {Alegre}, \citenamefont
  {Safavi-Naeini}, \citenamefont {Meenehan}, \citenamefont {Cohen},
  \citenamefont {Stobbe},\ and\ \citenamefont {Painter}}]{Winger.11.OP}%
  \BibitemOpen
  \bibfield  {author} {\bibinfo {author} {\bibfnamefont {M.}~\bibnamefont
  {Winger}}, \bibinfo {author} {\bibfnamefont {T.~D.}\ \bibnamefont {Blasius}},
  \bibinfo {author} {\bibfnamefont {T.~P.~Mayer}\ \bibnamefont {Alegre}},
  \bibinfo {author} {\bibfnamefont {A.~H.}\ \bibnamefont {Safavi-Naeini}},
  \bibinfo {author} {\bibfnamefont {S.}~\bibnamefont {Meenehan}}, \bibinfo
  {author} {\bibfnamefont {J.}~\bibnamefont {Cohen}}, \bibinfo {author}
  {\bibfnamefont {S.}~\bibnamefont {Stobbe}}, \ and\ \bibinfo {author}
  {\bibfnamefont {O.}~\bibnamefont {Painter}},\ }\bibfield  {title} {\enquote
  {\bibinfo {title} {A chip-scale integrated cavity-electro-optomechanics
  platform},}\ }\href {\doibase 10.1364/OE.19.024905} {\bibfield  {journal}
  {\bibinfo  {journal} {Opt. Express}\ }\textbf {\bibinfo {volume} {19}},\
  \bibinfo {pages} {24905--24921} (\bibinfo {year} {2011})}\BibitemShut
  {NoStop}%
\bibitem [{\citenamefont {Yeo}\ \emph {et~al.}(2014)\citenamefont {Yeo},
  \citenamefont {de~Assis}, \citenamefont {Gloppe}, \citenamefont
  {Dupont-Ferrier}, \citenamefont {Verlot}, \citenamefont {Malik},
  \citenamefont {Dupuy}, \citenamefont {Claudon}, \citenamefont {Gérard},
  \citenamefont {Auff\'eves}, \citenamefont {Nogues}, \citenamefont {Seidelin},
  \citenamefont {Poizat}, \citenamefont {Arcizet},\ and\ \citenamefont
  {Richard}}]{Yeo.14.NN}%
  \BibitemOpen
  \bibfield  {author} {\bibinfo {author} {\bibfnamefont {I.}~\bibnamefont
  {Yeo}}, \bibinfo {author} {\bibfnamefont {P-L.}\ \bibnamefont {de~Assis}},
  \bibinfo {author} {\bibfnamefont {A.}~\bibnamefont {Gloppe}}, \bibinfo
  {author} {\bibfnamefont {E.}~\bibnamefont {Dupont-Ferrier}}, \bibinfo
  {author} {\bibfnamefont {P.}~\bibnamefont {Verlot}}, \bibinfo {author}
  {\bibfnamefont {N.~S.}\ \bibnamefont {Malik}}, \bibinfo {author}
  {\bibfnamefont {E.}~\bibnamefont {Dupuy}}, \bibinfo {author} {\bibfnamefont
  {J.}~\bibnamefont {Claudon}}, \bibinfo {author} {\bibfnamefont {J-M.}\
  \bibnamefont {Gérard}}, \bibinfo {author} {\bibfnamefont {A.}~\bibnamefont
  {Auff\'eves}}, \bibinfo {author} {\bibfnamefont {G.}~\bibnamefont {Nogues}},
  \bibinfo {author} {\bibfnamefont {S.}~\bibnamefont {Seidelin}}, \bibinfo
  {author} {\bibfnamefont {J-Ph.}\ \bibnamefont {Poizat}}, \bibinfo {author}
  {\bibfnamefont {O.}~\bibnamefont {Arcizet}}, \ and\ \bibinfo {author}
  {\bibfnamefont {M.}~\bibnamefont {Richard}},\ }\bibfield  {title} {\enquote
  {\bibinfo {title} {Strain-mediated coupling in a quantum dot-mechanical
  oscillator hybrid system},}\ }\href {https://doi.org/10.1038/nnano.2013.274}
  {\bibfield  {journal} {\bibinfo  {journal} {Nat. Nanotechnol.}\ }\textbf
  {\bibinfo {volume} {9}},\ \bibinfo {pages} {106--110} (\bibinfo {year}
  {2014})}\BibitemShut {NoStop}%
\bibitem [{\citenamefont {Okazaki}\ \emph {et~al.}(2016)\citenamefont
  {Okazaki}, \citenamefont {Mahboob}, \citenamefont {Onomitsu}, \citenamefont
  {Sasaki},\ and\ \citenamefont {Yamaguchi}}]{Okazaki.16.NC}%
  \BibitemOpen
  \bibfield  {author} {\bibinfo {author} {\bibfnamefont {Y.}~\bibnamefont
  {Okazaki}}, \bibinfo {author} {\bibfnamefont {I.}~\bibnamefont {Mahboob}},
  \bibinfo {author} {\bibfnamefont {K.}~\bibnamefont {Onomitsu}}, \bibinfo
  {author} {\bibfnamefont {S.}~\bibnamefont {Sasaki}}, \ and\ \bibinfo {author}
  {\bibfnamefont {H.}~\bibnamefont {Yamaguchi}},\ }\bibfield  {title} {\enquote
  {\bibinfo {title} {Gate-controlled electromechanical backaction induced by a
  quantum dot},}\ }\href {https://doi.org/10.1038/ncomms11132} {\bibfield
  {journal} {\bibinfo  {journal} {Nat. Commun.}\ }\textbf {\bibinfo {volume}
  {7}},\ \bibinfo {pages} {11132--} (\bibinfo {year} {2016})}\BibitemShut
  {NoStop}%
\bibitem [{\citenamefont {Midolo}\ \emph {et~al.}(2018)\citenamefont {Midolo},
  \citenamefont {Schliesser},\ and\ \citenamefont {Fiore}}]{Midolo.18.NN}%
  \BibitemOpen
  \bibfield  {author} {\bibinfo {author} {\bibfnamefont {L.}~\bibnamefont
  {Midolo}}, \bibinfo {author} {\bibfnamefont {A.}~\bibnamefont {Schliesser}},
  \ and\ \bibinfo {author} {\bibfnamefont {A.}~\bibnamefont {Fiore}},\
  }\bibfield  {title} {\enquote {\bibinfo {title} {Nano-opto-electro-mechanical
  systems},}\ }\href {https://doi.org/10.1038/s41565-017-0039-1} {\bibfield
  {journal} {\bibinfo  {journal} {Nat. Nanotechnol.}\ }\textbf {\bibinfo
  {volume} {13}},\ \bibinfo {pages} {11} (\bibinfo {year} {2018})}\BibitemShut
  {NoStop}%
\bibitem [{\citenamefont {Wingreen}\ \emph {et~al.}(1989)\citenamefont
  {Wingreen}, \citenamefont {Jacobsen},\ and\ \citenamefont
  {Wilkins}}]{Wingreen.89.PRB}%
  \BibitemOpen
  \bibfield  {author} {\bibinfo {author} {\bibfnamefont {N.~S.}\ \bibnamefont
  {Wingreen}}, \bibinfo {author} {\bibfnamefont {K.~W.}\ \bibnamefont
  {Jacobsen}}, \ and\ \bibinfo {author} {\bibfnamefont {J.~W.}\ \bibnamefont
  {Wilkins}},\ }\bibfield  {title} {\enquote {\bibinfo {title} {Inelastic
  scattering in resonant tunneling},}\ }\href {\doibase
  10.1103/PhysRevB.40.11834} {\bibfield  {journal} {\bibinfo  {journal} {Phys.
  Rev. B}\ }\textbf {\bibinfo {volume} {40}},\ \bibinfo {pages} {11834--11850}
  (\bibinfo {year} {1989})}\BibitemShut {NoStop}%
\bibitem [{Note1()}]{Note1}%
  \BibitemOpen
  \bibinfo {note} {We point out that Eq.~\protect \textup {\hbox {\mathsurround
  \z@ \protect \normalfont (\ignorespaces \ref {h0}\unskip \@@italiccorr )}} is
  applicable in the sequential tunneling regime regardless of whether the
  solid-state SET is superconducting or not \cite
  {Clerk.05.NJP,Rodrigues.05.NJP}. The only difference is that in the case of
  an superconducting SET, one should interpret $d^{\protect \dagger }d$ as the
  occupation operator for quasi-particles. For our case, although we are in the
  superconducting regime ($T\sim 100$ mK), the normal-state description is
  still applicable as one can apply an out-of-plane magnetic field to turn a
  superconducting SET to a normal one \cite {Knobel.03.N}.}\BibitemShut {Stop}%
\bibitem [{SM.()}]{SM.20.NULL}%
  \BibitemOpen
  \href@noop {} {}\bibinfo {note} {See Supplemental Material for the
  implementation of the generalized input-output method and details related to
  the derivations of current-voltage characteristics of the single electron
  transistor.}\BibitemShut {Stop}%
\bibitem [{\citenamefont {Liu}\ and\ \citenamefont
  {Segal}(2020{\natexlab{a}})}]{Liu.20.PRB}%
  \BibitemOpen
  \bibfield  {author} {\bibinfo {author} {\bibfnamefont {J.}~\bibnamefont
  {Liu}}\ and\ \bibinfo {author} {\bibfnamefont {D.}~\bibnamefont {Segal}},\
  }\bibfield  {title} {\enquote {\bibinfo {title} {Generalized input-output
  method to quantum transport junctions. i. general formulation},}\ }\href
  {\doibase 10.1103/PhysRevB.101.155406} {\bibfield  {journal} {\bibinfo
  {journal} {Phys. Rev. B}\ }\textbf {\bibinfo {volume} {101}},\ \bibinfo
  {pages} {155406} (\bibinfo {year} {2020}{\natexlab{a}})}\BibitemShut
  {NoStop}%
\bibitem [{\citenamefont {Liu}\ and\ \citenamefont
  {Segal}(2020{\natexlab{b}})}]{Liu.20.PRBa}%
  \BibitemOpen
  \bibfield  {author} {\bibinfo {author} {\bibfnamefont {J.}~\bibnamefont
  {Liu}}\ and\ \bibinfo {author} {\bibfnamefont {D.}~\bibnamefont {Segal}},\
  }\bibfield  {title} {\enquote {\bibinfo {title} {Generalized input-output
  method to quantum transport junctions. ii. applications},}\ }\href {\doibase
  10.1103/PhysRevB.101.155407} {\bibfield  {journal} {\bibinfo  {journal}
  {Phys. Rev. B}\ }\textbf {\bibinfo {volume} {101}},\ \bibinfo {pages}
  {155407} (\bibinfo {year} {2020}{\natexlab{b}})}\BibitemShut {NoStop}%
\bibitem [{\citenamefont {Armour}\ \emph {et~al.}(2004)\citenamefont {Armour},
  \citenamefont {Blencowe},\ and\ \citenamefont {Zhang}}]{Armour.04.PRB}%
  \BibitemOpen
  \bibfield  {author} {\bibinfo {author} {\bibfnamefont {A.~D.}\ \bibnamefont
  {Armour}}, \bibinfo {author} {\bibfnamefont {M.~P.}\ \bibnamefont
  {Blencowe}}, \ and\ \bibinfo {author} {\bibfnamefont {Y.}~\bibnamefont
  {Zhang}},\ }\bibfield  {title} {\enquote {\bibinfo {title} {Classical
  dynamics of a nanomechanical resonator coupled to a single-electron
  transistor},}\ }\href {\doibase 10.1103/PhysRevB.69.125313} {\bibfield
  {journal} {\bibinfo  {journal} {Phys. Rev. B}\ }\textbf {\bibinfo {volume}
  {69}},\ \bibinfo {pages} {125313} (\bibinfo {year} {2004})}\BibitemShut
  {NoStop}%
\bibitem [{\citenamefont {Clerk}(2004)}]{Clerk.04.PRB}%
  \BibitemOpen
  \bibfield  {author} {\bibinfo {author} {\bibfnamefont {A.~A.}\ \bibnamefont
  {Clerk}},\ }\bibfield  {title} {\enquote {\bibinfo {title} {Quantum-limited
  position detection and amplification: A linear response perspective},}\
  }\href {\doibase 10.1103/PhysRevB.70.245306} {\bibfield  {journal} {\bibinfo
  {journal} {Phys. Rev. B}\ }\textbf {\bibinfo {volume} {70}},\ \bibinfo
  {pages} {245306} (\bibinfo {year} {2004})}\BibitemShut {NoStop}%
\bibitem [{\citenamefont {Clerk}\ and\ \citenamefont
  {Bennett}(2005)}]{Clerk.05.NJP}%
  \BibitemOpen
  \bibfield  {author} {\bibinfo {author} {\bibfnamefont {A.~A.}\ \bibnamefont
  {Clerk}}\ and\ \bibinfo {author} {\bibfnamefont {S.}~\bibnamefont
  {Bennett}},\ }\bibfield  {title} {\enquote {\bibinfo {title} {Quantum
  nanoelectromechanics with electrons, quasi-particles and cooper pairs:
  effective bath descriptions and strong feedback effects},}\ }\href {\doibase
  10.1088/1367-2630/7/1/238} {\bibfield  {journal} {\bibinfo  {journal} {New J.
  Phys.}\ }\textbf {\bibinfo {volume} {7}},\ \bibinfo {pages} {238--238}
  (\bibinfo {year} {2005})}\BibitemShut {NoStop}%
\bibitem [{\citenamefont {Rodrigues}\ and\ \citenamefont
  {Armour}(2005)}]{Rodrigues.05.NJP}%
  \BibitemOpen
  \bibfield  {author} {\bibinfo {author} {\bibfnamefont {D.~A.}\ \bibnamefont
  {Rodrigues}}\ and\ \bibinfo {author} {\bibfnamefont {A.~D.}\ \bibnamefont
  {Armour}},\ }\bibfield  {title} {\enquote {\bibinfo {title} {Quantum master
  equation descriptions of a nanomechanical resonator coupled to a
  single-electron transistor},}\ }\href {\doibase 10.1088/1367-2630/7/1/251}
  {\bibfield  {journal} {\bibinfo  {journal} {New J. Phys.}\ }\textbf {\bibinfo
  {volume} {7}},\ \bibinfo {pages} {251--251} (\bibinfo {year}
  {2005})}\BibitemShut {NoStop}%
\bibitem [{\citenamefont {Bennett}\ \emph {et~al.}(2010)\citenamefont
  {Bennett}, \citenamefont {Cockins}, \citenamefont {Miyahara}, \citenamefont
  {Gr\"utter},\ and\ \citenamefont {Clerk}}]{Bennett.10.PRL}%
  \BibitemOpen
  \bibfield  {author} {\bibinfo {author} {\bibfnamefont {S.~D.}\ \bibnamefont
  {Bennett}}, \bibinfo {author} {\bibfnamefont {L.}~\bibnamefont {Cockins}},
  \bibinfo {author} {\bibfnamefont {Y.}~\bibnamefont {Miyahara}}, \bibinfo
  {author} {\bibfnamefont {P.}~\bibnamefont {Gr\"utter}}, \ and\ \bibinfo
  {author} {\bibfnamefont {A.~A.}\ \bibnamefont {Clerk}},\ }\bibfield  {title}
  {\enquote {\bibinfo {title} {Strong electromechanical coupling of an atomic
  force microscope cantilever to a quantum dot},}\ }\href {\doibase
  10.1103/PhysRevLett.104.017203} {\bibfield  {journal} {\bibinfo  {journal}
  {Phys. Rev. Lett.}\ }\textbf {\bibinfo {volume} {104}},\ \bibinfo {pages}
  {017203} (\bibinfo {year} {2010})}\BibitemShut {NoStop}%
\bibitem [{\citenamefont {Khivrich}\ \emph {et~al.}(2019)\citenamefont
  {Khivrich}, \citenamefont {Clerk},\ and\ \citenamefont
  {Ilani}}]{Khivrich.19.NN}%
  \BibitemOpen
  \bibfield  {author} {\bibinfo {author} {\bibfnamefont {I.}~\bibnamefont
  {Khivrich}}, \bibinfo {author} {\bibfnamefont {A.~A.}\ \bibnamefont {Clerk}},
  \ and\ \bibinfo {author} {\bibfnamefont {S.}~\bibnamefont {Ilani}},\
  }\bibfield  {title} {\enquote {\bibinfo {title} {Nanomechanical pump-probe
  measurements of insulating electronic states in a carbon nanotube},}\ }\href
  {https://doi.org/10.1038/s41565-018-0341-6} {\bibfield  {journal} {\bibinfo
  {journal} {Nat. Nanotechnol.}\ }\textbf {\bibinfo {volume} {14}},\ \bibinfo
  {pages} {161--167} (\bibinfo {year} {2019})}\BibitemShut {NoStop}%
\bibitem [{\citenamefont {Devoret}\ and\ \citenamefont
  {Schoelkopf}(2000)}]{Devoret.00.N}%
  \BibitemOpen
  \bibfield  {author} {\bibinfo {author} {\bibfnamefont {M.~H.}\ \bibnamefont
  {Devoret}}\ and\ \bibinfo {author} {\bibfnamefont {R.~J.}\ \bibnamefont
  {Schoelkopf}},\ }\bibfield  {title} {\enquote {\bibinfo {title} {Amplifying
  quantum signals with the single-electron transistor},}\ }\href
  {https://doi.org/10.1038/35023253} {\bibfield  {journal} {\bibinfo  {journal}
  {Nature}\ }\textbf {\bibinfo {volume} {406}},\ \bibinfo {pages} {1039--1046}
  (\bibinfo {year} {2000})}\BibitemShut {NoStop}%
\bibitem [{\citenamefont {Ouyang}\ \emph {et~al.}(2009)\citenamefont {Ouyang},
  \citenamefont {You},\ and\ \citenamefont {Nori}}]{Ouyang.09.PRB}%
  \BibitemOpen
  \bibfield  {author} {\bibinfo {author} {\bibfnamefont {S-H}\ \bibnamefont
  {Ouyang}}, \bibinfo {author} {\bibfnamefont {J.~Q.}\ \bibnamefont {You}}, \
  and\ \bibinfo {author} {\bibfnamefont {F.}~\bibnamefont {Nori}},\ }\bibfield
  {title} {\enquote {\bibinfo {title} {Cooling a mechanical resonator via
  coupling to a tunable double quantum dot},}\ }\href {\doibase
  10.1103/PhysRevB.79.075304} {\bibfield  {journal} {\bibinfo  {journal} {Phys.
  Rev. B}\ }\textbf {\bibinfo {volume} {79}},\ \bibinfo {pages} {075304}
  (\bibinfo {year} {2009})}\BibitemShut {NoStop}%
\bibitem [{\citenamefont {Liu}\ \emph {et~al.}(2018)\citenamefont {Liu},
  \citenamefont {Wang}, \citenamefont {Zhang},\ and\ \citenamefont
  {Liu}}]{Liu.18.CPB}%
  \BibitemOpen
  \bibfield  {author} {\bibinfo {author} {\bibfnamefont {Y.}~\bibnamefont
  {Liu}}, \bibinfo {author} {\bibfnamefont {C.}~\bibnamefont {Wang}}, \bibinfo
  {author} {\bibfnamefont {J.}~\bibnamefont {Zhang}}, \ and\ \bibinfo {author}
  {\bibfnamefont {Y.}~\bibnamefont {Liu}},\ }\bibfield  {title} {\enquote
  {\bibinfo {title} {Cavity optomechanics: Manipulating photons and phonons
  towards the single-photon strong coupling},}\ }\href {\doibase
  10.1088/1674-1056/27/2/024204} {\bibfield  {journal} {\bibinfo  {journal}
  {Chin. Phys. B}\ }\textbf {\bibinfo {volume} {27}},\ \bibinfo {pages}
  {024204} (\bibinfo {year} {2018})}\BibitemShut {NoStop}%
\bibitem [{\citenamefont {Brennecke}\ \emph {et~al.}(2008)\citenamefont
  {Brennecke}, \citenamefont {Ritter}, \citenamefont {Donner},\ and\
  \citenamefont {Esslinger}}]{Brennecke.S.08}%
  \BibitemOpen
  \bibfield  {author} {\bibinfo {author} {\bibfnamefont {F?}\ \bibnamefont
  {Brennecke}}, \bibinfo {author} {\bibfnamefont {S?}\ \bibnamefont
  {Ritter}}, \bibinfo {author} {\bibfnamefont {T?}\ \bibnamefont {Donner}}, \
  and\ \bibinfo {author} {\bibfnamefont {T?}\ \bibnamefont {Esslinger}},\
  }\bibfield  {title} {\enquote {\bibinfo {title} {Cavity optomechanics with a
  bose-einstein condensate},}\ }\href {\doibase 10.1126/science.1163218}
  {\bibfield  {journal} {\bibinfo  {journal} {Science}\ }\textbf {\bibinfo
  {volume} {322}},\ \bibinfo {pages} {235--238} (\bibinfo {year}
  {2008})}\BibitemShut {NoStop}%
\bibitem [{\citenamefont {Murch}\ \emph {et~al.}(2008)\citenamefont {Murch},
  \citenamefont {Moore}, \citenamefont {Gupta},\ and\ \citenamefont
  {Stamper-Kurn}}]{Murch.08.NP}%
  \BibitemOpen
  \bibfield  {author} {\bibinfo {author} {\bibfnamefont {K.~W.}\ \bibnamefont
  {Murch}}, \bibinfo {author} {\bibfnamefont {K.~L.}\ \bibnamefont {Moore}},
  \bibinfo {author} {\bibfnamefont {S.}~\bibnamefont {Gupta}}, \ and\ \bibinfo
  {author} {\bibfnamefont {D.~M.}\ \bibnamefont {Stamper-Kurn}},\ }\bibfield
  {title} {\enquote {\bibinfo {title} {Observation of quantum-measurement
  backaction with an ultracold atomic gas},}\ }\href
  {https://doi.org/10.1038/nphys965} {\bibfield  {journal} {\bibinfo  {journal}
  {Nat. Phys.}\ }\textbf {\bibinfo {volume} {4}},\ \bibinfo {pages} {561--564}
  (\bibinfo {year} {2008})}\BibitemShut {NoStop}%
\bibitem [{\citenamefont {Davan\c{c}o}\ \emph {et~al.}(2012)\citenamefont
  {Davan\c{c}o}, \citenamefont {Chan}, \citenamefont {Safavi-Naeini},
  \citenamefont {Painter},\ and\ \citenamefont {Srinivasan}}]{Davanco.12.OP}%
  \BibitemOpen
  \bibfield  {author} {\bibinfo {author} {\bibfnamefont {M.}~\bibnamefont
  {Davan\c{c}o}}, \bibinfo {author} {\bibfnamefont {J.}~\bibnamefont {Chan}},
  \bibinfo {author} {\bibfnamefont {A.~H.}\ \bibnamefont {Safavi-Naeini}},
  \bibinfo {author} {\bibfnamefont {O.}~\bibnamefont {Painter}}, \ and\
  \bibinfo {author} {\bibfnamefont {K.}~\bibnamefont {Srinivasan}},\ }\bibfield
   {title} {\enquote {\bibinfo {title} {Slot-mode-coupled optomechanical
  crystals},}\ }\href {\doibase 10.1364/OE.20.024394} {\bibfield  {journal}
  {\bibinfo  {journal} {Opt. Express}\ }\textbf {\bibinfo {volume} {20}},\
  \bibinfo {pages} {24394--24410} (\bibinfo {year} {2012})}\BibitemShut
  {NoStop}%
\bibitem [{\citenamefont {Zoepfl}\ \emph {et~al.}(2020)\citenamefont {Zoepfl},
  \citenamefont {Juan}, \citenamefont {Schneider},\ and\ \citenamefont
  {Kirchmair}}]{Zoepfl.20.PRL}%
  \BibitemOpen
  \bibfield  {author} {\bibinfo {author} {\bibfnamefont {D.}~\bibnamefont
  {Zoepfl}}, \bibinfo {author} {\bibfnamefont {M.~L.}\ \bibnamefont {Juan}},
  \bibinfo {author} {\bibfnamefont {C.~M.~F.}\ \bibnamefont {Schneider}}, \
  and\ \bibinfo {author} {\bibfnamefont {G.}~\bibnamefont {Kirchmair}},\
  }\bibfield  {title} {\enquote {\bibinfo {title} {Single-photon cooling in
  microwave magnetomechanics},}\ }\href {\doibase
  10.1103/PhysRevLett.125.023601} {\bibfield  {journal} {\bibinfo  {journal}
  {Phys. Rev. Lett.}\ }\textbf {\bibinfo {volume} {125}},\ \bibinfo {pages}
  {023601} (\bibinfo {year} {2020})}\BibitemShut {NoStop}%
\bibitem [{\citenamefont {Ares}\ \emph {et~al.}(2016)\citenamefont {Ares},
  \citenamefont {Pei}, \citenamefont {Mavalankar}, \citenamefont
  {Mergenthaler}, \citenamefont {Warner}, \citenamefont {Briggs},\ and\
  \citenamefont {Laird}}]{Ares.16.PRL}%
  \BibitemOpen
  \bibfield  {author} {\bibinfo {author} {\bibfnamefont {N.}~\bibnamefont
  {Ares}}, \bibinfo {author} {\bibfnamefont {T.}~\bibnamefont {Pei}}, \bibinfo
  {author} {\bibfnamefont {A.}~\bibnamefont {Mavalankar}}, \bibinfo {author}
  {\bibfnamefont {M.}~\bibnamefont {Mergenthaler}}, \bibinfo {author}
  {\bibfnamefont {J.~H.}\ \bibnamefont {Warner}}, \bibinfo {author}
  {\bibfnamefont {G.~A.~D.}\ \bibnamefont {Briggs}}, \ and\ \bibinfo {author}
  {\bibfnamefont {E.~A.}\ \bibnamefont {Laird}},\ }\bibfield  {title} {\enquote
  {\bibinfo {title} {Resonant optomechanics with a vibrating carbon nanotube
  and a radio-frequency cavity},}\ }\href {\doibase
  10.1103/PhysRevLett.117.170801} {\bibfield  {journal} {\bibinfo  {journal}
  {Phys. Rev. Lett.}\ }\textbf {\bibinfo {volume} {117}},\ \bibinfo {pages}
  {170801} (\bibinfo {year} {2016})}\BibitemShut {NoStop}%
\bibitem [{\citenamefont {Frisk~Kockum}\ \emph {et~al.}(2019)\citenamefont
  {Frisk~Kockum}, \citenamefont {Miranowicz}, \citenamefont {De~Liberato},
  \citenamefont {Savasta},\ and\ \citenamefont
  {Nori}}]{frisk_kockum_ultrastrong_2019}%
  \BibitemOpen
  \bibfield  {author} {\bibinfo {author} {\bibfnamefont {A.}~\bibnamefont
  {Frisk~Kockum}}, \bibinfo {author} {\bibfnamefont {A.}~\bibnamefont
  {Miranowicz}}, \bibinfo {author} {\bibfnamefont {S.}~\bibnamefont
  {De~Liberato}}, \bibinfo {author} {\bibfnamefont {S.}~\bibnamefont
  {Savasta}}, \ and\ \bibinfo {author} {\bibfnamefont {F.}~\bibnamefont
  {Nori}},\ }\bibfield  {title} {\enquote {\bibinfo {title} {Ultrastrong
  coupling between light and matter},}\ }\href {\doibase
  10.1038/s42254-018-0006-2} {\bibfield  {journal} {\bibinfo  {journal} {Nat.
  Rev. Phys.}\ }\textbf {\bibinfo {volume} {1}},\ \bibinfo {pages} {19--40}
  (\bibinfo {year} {2019})}\BibitemShut {NoStop}%
\bibitem [{\citenamefont {Press}\ \emph {et~al.}(2007)\citenamefont {Press},
  \citenamefont {G{\"o}tzinger}, \citenamefont {Reitzenstein}, \citenamefont
  {Hofmann}, \citenamefont {L{\"o}ffler}, \citenamefont {Kamp}, \citenamefont
  {Forchel},\ and\ \citenamefont {Yamamoto}}]{press2007photon}%
  \BibitemOpen
  \bibfield  {author} {\bibinfo {author} {\bibfnamefont {D.}~\bibnamefont
  {Press}}, \bibinfo {author} {\bibfnamefont {S.}~\bibnamefont
  {G{\"o}tzinger}}, \bibinfo {author} {\bibfnamefont {S.}~\bibnamefont
  {Reitzenstein}}, \bibinfo {author} {\bibfnamefont {C.}~\bibnamefont
  {Hofmann}}, \bibinfo {author} {\bibfnamefont {A.}~\bibnamefont
  {L{\"o}ffler}}, \bibinfo {author} {\bibfnamefont {M.}~\bibnamefont {Kamp}},
  \bibinfo {author} {\bibfnamefont {A.}~\bibnamefont {Forchel}}, \ and\
  \bibinfo {author} {\bibfnamefont {Y.}~\bibnamefont {Yamamoto}},\ }\bibfield
  {title} {\enquote {\bibinfo {title} {Photon antibunching from a single
  quantum-dot-microcavity system in the strong coupling regime},}\ }\href
  {\doibase 10.1103/PhysRevLett.98.117402} {\bibfield  {journal} {\bibinfo
  {journal} {Phys. Rev. Lett.}\ }\textbf {\bibinfo {volume} {98}},\ \bibinfo
  {pages} {117402} (\bibinfo {year} {2007})}\BibitemShut {NoStop}%
\end{thebibliography}

%

\clearpage
\renewcommand{\thesection}{\Roman{section}} 
\renewcommand{\thesubsection}{\Alph{subsection}}
\renewcommand{\theequation}{S\arabic{equation}}
\renewcommand{\thefigure}{S\arabic{figure}}
\renewcommand{\thetable}{S\arabic{table}}
\setcounter{equation}{0}  
\setcounter{figure}{0}

\begin{widetext}

{\Large{\bf Supplemental material:} Quantum Nondemolition Photon Counting With a Hybrid Electromechanical Probe}
\\
\\
\\
In this supplementary material we present the derivation of the steady state charge current expression used in the main text by resorting to a generalized input-output method \cite{Liu.20.PRB,Liu.20.PRBa}.

\section{I. Current-voltage characteristics of single electron transistors}
\label{a:1}
In this study, we consider a hybrid optoelectromechanical system, which includes a quantum cavity coupled to a mechanical mode, itself interacting with a single electron transistor (SET) that acts as an electromechanical probe to the photon number. For the sake of completeness, we first write down the total Hamiltonian $H_{\mathrm{tot}}=H_0+H_E+H_{\mathrm{diss}}$ (setting $\hbar=1$, $e=1$, $k_B=1$ and Fermi energy $\epsilon_F=0$ hereafter),
\bea\label{hs_a}
H_0 &=& \epsilon_0d^{\dagger}d+\omega_c a^{\dagger}a+\omega_bb^{\dagger}b -g_0a^{\dagger}a(b^{\dagger}+b)+\lambda d^{\dagger}d(b^{\dagger}+b), \nonumber\\
H_E &=& \sum_{k,v=S,D}\Big[\epsilon_{kv}c_{kv}^{\dagger}c_{kv}+t_{kv}(c_{kv}^{\dagger}d+d^{\dagger}c_{kv})\Big],\nonumber\\
H_{\mathrm{diss}} &=& \sum_{j}\omega_{j}r_{j}^{\dagger}r_{j}+ \sum_{j}\eta_{j}(r_{j}^{\dagger}b+b^{\dagger}r_{j}).
\eea
Here, $H_0$ accounts for the high-quality single mode cavity of frequency $\omega_c$ with an annihilation operator $a$, the high-quality mechanical oscillator of frequency $\omega_b$ with an annihilation operator $b$, a SET conductor of an electrostatic energy $\epsilon_0$ with an annihilation operator $d$, and a radiation-pressure optomechanical coupling as well as an electromechanical interaction characterized by coupling strengths $g_0$ and $\lambda$, respectively. $H_E$ contains the electron source (S) and drain (D) of the SET, together with electron tunneling between the conductor and the electrodes. Finally, $H_{\mathrm{diss}}$ accounts for the damping of the mechanical mode induced by its intrinsic thermal environment modelled as a harmonic thermal bath with annihilation operators $r_{j}$ and 
frequencies $\omega_{j}$. $\eta_j$ denotes the coupling strength between the mechanical mode and the $j$th harmonic oscillator of the thermal bath. We assume that the interaction between the mechanical mode and the thermalized modes is rather weak 
such that the rotating wave approximation is justified. 
The influence of thermal bath, acting on the mechanical mode, is characterized by the spectral density function $\gamma_0(\omega)=\pi\sum_j\eta_{j}^2\delta(\omega-\omega_{j})$. 

The transformed Hamiltonian $\tilde{H}_{\mathrm{tot}}\equiv\mathcal{G}H_{\mathrm{tot}}\mathcal{G}^{\dagger}=\tilde{H}_0+\tilde{H}_E+H_{\mathrm{diss}}$ under a unitary transformation generated by the operator $\mathcal{G}=\exp\Big[-g_0(b^{\dagger}-b)a^{\dagger}a/\omega_b\Big]\otimes\exp\Big[\lambda(b^{\dagger}-b)d^{\dagger}d/\omega_b\Big]$ becomes
\bea
&&\tilde{H}_0~=~\left(\epsilon_0-\frac{\lambda^2}{\omega_b}\right)d^{\dagger}d+\omega_c a^{\dagger}a+\omega_bb^{\dagger}b-\frac{g_0^2}{\omega_b}(a^{\dagger}a)^2+\frac{2\lambda g_0}{\omega_b}a^{\dagger}ad^{\dagger}d,\nonumber\\
&&\tilde{H}_E~=~\sum_{k,v=S,D}\Big[\epsilon_{kv}c_{kv}^{\dagger}c_{kv}+t_{kv}(c_{kv}^{\dagger}\tilde{d}+\tilde{d}^{\dagger}c_{kv})\Big].
\eea
Here $\tilde{d}\equiv\mathcal{D}_{\lambda}^{\dagger}d$ denotes a polaron operator with a displacement operator defined as $\mathcal{D}_{\lambda}=\exp[(b^{\dagger}-b)\lambda/\omega_b]$. We neglect the effect of this transformation on $H_{\mathrm{diss}}$. To be precise, we ignore the term 
 $\sum_{j}\eta_{j}(\lambda d^{\dagger}d/\omega_b-g_0 a^{\dagger}a/\omega_b) (r_{j}^{\dagger}+r_{j})$ in the transformed Hamiltonian. 
This omission is justified in the present study since the energies 
$\eta_{j}\lambda/\omega_b$ and $\eta_j g_0/\omega_b$ are assumed small, by noting that the coupling between the high-quality mechanical mode and its thermal environment should be rather weak and $\lambda,g_0\ll\omega_b$.

Adopting a recently developed generalized input-output method for electronic systems \cite{Liu.20.PRB,Liu.20.PRBa}, we treat the hybrid quantum system within a unified input-output picture. As the system $\tilde{H}_0$ contains both fermionic and bosonic operators, we should treat them separately. To this end, we use the notations 
$[A,B]\equiv[A,B]_-$ and $\{A,B\}\equiv[A,B]_+$ for the quantum commutator and anti-commutator, respectively. The corresponding Heisenberg-Langevin equation (HLE) that governs the dynamical evolution of system operators reads \cite{Liu.20.PRB}
\begin{equation}
\label{eq:eom_o}
\dot{\mathcal{O}}~=~ i[\tilde{H}_{0},\mathcal{O}]_{-}-i\sum_{v=S,D}\mathbb{L}_{\pm}^v-i\mathbb{X}.
\end{equation}
Here, $\dot{A}$ denotes a time derivative of operator $A$. Explicit forms for the superoperators $\mathbb{L}_{\pm}^v$ and $\mathbb{X}$ are obtained from an input-output description of the electron tunneling Hamiltonian in $\tilde{H}_E$ and the thermal damping of the mechanical mode by $H_{\mathrm{diss}}$, respectively \cite{Liu.20.PRB},
\bea
\mathbb{L}_{\pm}^v&\equiv& \mp\left(i\Gamma_v \tilde{d}^{\dagger}+\sqrt{2\pi}d_{in}^{v,\dagger}\right)[\mathcal{O},\tilde{d}]_{\pm}+[\mathcal{O},\tilde{d}^{\dagger}]_{\pm}\left(-i\Gamma_v\tilde{d}+\sqrt{2\pi}d_{in}^v\right), \nonumber\\
\mathbb{X} &\equiv & \left(i\gamma_0 b^{\dagger}+\sqrt{2\pi}b_{0,in}^{\dagger}\right)[\mathcal{O},b]_{-}+[\mathcal{O},b^{\dagger}]_-\left(-i\gamma_0 b+\sqrt{2\pi}b_{0,in}\right).
\eea
Here $\gamma_0\equiv\gamma_0(\omega_{b})$ denotes a damping rate for the mechanical mode induced by its thermal bath,
$\Gamma_v(\epsilon)=\pi\sum_kt_{kv}^2\delta(\epsilon-\epsilon_{kv})$ is the spectral density function of electrons in the two metals.
In the above equation,
 the top signs apply if $\mathcal{O}$ is a fermionic operator; the bottom signs apply if $\mathcal{O}$ is bosonic. We remark that the form of $\mathbb{L}_{\pm}^v$ is exact in the wide-band limit, whereas $\mathbb{X}$ is obtained by assuming a Markovian thermal bath with $\gamma_0(\omega)$ assumed a constant at the vicinity of $\omega_{b}$. We have defined input fields as follows
\bea
d_{in}^v(t) &\equiv& \frac{1}{\sqrt{2\pi}} \sum_kt_{kv}e^{-i\epsilon_{kv}(t-t_0)}c_{kv}(t_0),\nonumber\\
b_{0,in}(t) &\equiv& \frac{1}{\sqrt{2\pi}} \sum_j\eta_{j}e^{-i\omega_{j}(t-t_0)}r_{j}(t_0).
\eea 
Here $t_0$ is the initial time at which the dynamical evolution begins.

As can be seen, the definitions of input fields in terms of environment operators at the initial time
ensure that they can be specified as initial conditions. 
We prepare the initial state of the hybrid system to be such that, 
at $t=t_0$, the SET island, the mechanical mode, the cavity and their environments are decoupled. 
Specifically, we assume that the metal leads and the mechanical thermal bath are initially in their thermal 
equilibrium states characterized by the Fermi-Dirac distribution function
$n_F^v(\epsilon)=\{\exp[(\epsilon-\mu_v)/T_0]+1\}^{-1}$ with 
$\mu_v$ the chemical potentials and $T_0$ the ambient temperature, 
and the Bose-Einstein distribution function $n_{b,0}(\omega)=[\exp(\omega/T_0)-1]^{-1}$, respectively. We assume that the intrinsic thermal environment of the mechanical mode and metallic leads have the same ambient temperature $T_0$.
By doing so, the noise correlators associated with the input fields are given by \cite{Liu.20.PRB}
\bea\label{eq:corr_input}
\langle b_{0,in}^{\dagger}(t')b_{0,in}(t)\rangle &=& \gamma_0\int \frac{d\omega}{2\pi^2}e^{-i\omega(t-t')}n_{b,0}(\omega),\nonumber\\
\langle b_{0,in}(t)b_{0,in}^{\dagger}(t')\rangle &=& \gamma_0\int \frac{d\omega}{2\pi^2}e^{-i\omega(t-t')}[1+n_{b,0}(\omega)],\nonumber\\
\langle d_{in}^{v,\dagger}(t')d_{in}^{v'}(t)\rangle &=& \delta_{vv'}\Gamma_v\int \frac{d\epsilon}{2\pi^2}e^{-i\epsilon(t-t')}n_F^v(\epsilon),\nonumber\\
\langle d_{in}^v(t)d_{in}^{v',\dagger}(t')\rangle &=& \delta_{vv'}\Gamma_v\int \frac{d\epsilon}{2\pi^2}e^{-i\epsilon(t-t')}\left[1-n_F^v(\epsilon)\right].
\eea
In obtaining the first two correlation functions, we have approximated $\gamma_0(\omega)\simeq\gamma_0$, which is valid in the Markovian limit. The output fields are related to the input fields via the so-called input-output relations
\bea
b_{0,out}(t) &=& b_{0,in}(t)-i\sqrt{\frac{2}{\pi}}\gamma_0 b(t),\nonumber\\
d_{out}^v(t) &=& d_{in}^v(t)-i\sqrt{\frac{2}{\pi}}\Gamma_v\tilde{d}(t).
\eea
The above relations imply that it is sufficient to work with input fields in the context of input-output theory.

As the photon occupation $\bar{n}_p\equiv\langle a^{\dagger}a\rangle$ is time independent during the charge current measurement, we focus here on the dynamical evolution of electron and mechanical mode. Using the HLE Eq. (\ref{eq:eom_o}), we first find
\bea\label{eq:dot_b}
\dot{b}(t) &=& -\left(i\omega_b+\gamma_0\right)b(t)-i\sqrt{2\pi}b_{0,in}(t)+i\frac{\lambda}{\omega_b}\sum_v\left(i\Gamma_v\tilde{d}^{\dagger}(t)+\sqrt{2\pi}d_{in}^{v,\dagger}(t)\right)\tilde{d}(t)\nonumber\\
&&-i\frac{\lambda}{\omega_b}\sum_v\tilde{d}^{\dagger}(t)\left(-i\Gamma_v\tilde{d}(t)+\sqrt{2\pi}d_{in}^v(t)\right)\nonumber\\
&=& -\left(i\omega_b+\gamma_0\right)b(t)-i\sqrt{2\pi}b_{0,in}(t)+\frac{\lambda}{\omega_b}\sum_v\left[2\sqrt{2\pi}\mathrm{Im}\left(\tilde{d}^{\dagger}(t)d_{in}^v(t)\right)-2\Gamma_vd^{\dagger}(t)d(t)\right]\nonumber\\
&=& -\left(i\omega_b+\gamma_0\right)b(t)-i\sqrt{2\pi}b_{0,in}(t)+\frac{\lambda}{\omega_b}\sum_vJ_v(t),
\eea
where we have utilized the relations $[b,\mathcal{D}_{\lambda}^{\dagger}]=-\frac{\lambda}{\omega_b}\mathcal{D}_{\lambda}^{\dagger}$ and $[b,\mathcal{D}_{\lambda}]=\frac{\lambda}{\omega_b}\mathcal{D}_{\lambda}$. `Im' takes the imaginary part. $J_v$ is the formal definition of charge current operator out of $v$-lead \cite{Liu.20.PRB}
\begin{equation}\label{eq:jl_a}
J_v~=~2\sqrt{2\pi}\mathrm{Im}\left(\tilde{d}^{\dagger}d_{in}^v\right)-2\Gamma_vd^{\dagger}d. 
\end{equation}
Clearly, the term $\frac{\lambda}{\omega_b}\sum_vJ_v(t)$ in Eq. (\ref{eq:jl_a}) represents the backaction from the conducting electrons arising due to the coupling of the mechanical mode to the SET. In the steady state limit, we have $\sum_vJ_v=0$ because of charge conservation. However, at transient times, $\sum_vJ_v(t)$ is generally nonzero. To account for this dissipation source which will in turn affect the current-voltage characteristics of the SET through the mechanical backaction, we need a faithful treatment of backaction from the conducting electrons.

Technically speaking, this coupled dynamical problem is challenging to solve even numerically. To simplify the problem while taking into account the backactions, we resort to an effective treatment motivated by a significant time-scale separation between electron tunneling and mechanical motion as $\Gamma_v\gg\omega_b$ \cite{Clerk.04.PRB,Armour.04.PRB,Blencowe.04.PR,Clerk.05.NJP,Rodrigues.05.NJP,Naik.06.N,Blien.20.NC}: For a slow mechanical motion, an adiabatic approximation is valid and the SET acts as a thermalized environment characterized by a temperature $T_1$, and it induces an extra damping rate $\gamma_1$ on the mechanical mode. Particularly, $T_1$ is set by the source-drain voltage bias \cite{Armour.04.PRB,Clerk.05.NJP,Rodrigues.05.NJP} and $\gamma_1/\gamma_0\sim20-50$ \cite{Naik.06.N,Bennett.10.PRL}. Altogether, the mechanical mode experiences damping due to its direct thermal bath (temperature $T_0$ and decay rate $\gamma_0$) and from the electronic compartment (temperature $T_1$ and decay rate $\gamma_1$). These two processes sum up to a total effective damping with an effective damping rate $\gamma_{\mathrm{eff}}\equiv\gamma_0+\gamma_1$ and an effective temperature $T_{\mathrm{eff}}=(\gamma_0T_0+\gamma_1T_1)/\gamma_{\mathrm{eff}}$ \cite{Clerk.04.PRB,Clerk.05.NJP}. In doing so, the effective equation of motion for $b$ becomes
\begin{equation}\label{eq:b_effective}
\dot{b}~\simeq~-\left(i\omega_b+\gamma_{\mathrm{eff}}\right)b(t)-i\sqrt{2\pi}\tilde{b}_{in,\mathrm{eff}}(t),
\end{equation}
where the effective input field is determined by the following correlation functions 
\bea\label{eq:bin_corr}
\langle \tilde{b}_{in,\mathrm{eff}}^{\dagger}(t')\tilde{b}_{in,\mathrm{eff}}(t)\rangle &=& \gamma_{\mathrm{eff}}\int \frac{d\omega}{2\pi^2}e^{-i\omega(t-t')}n_{b,\mathrm{eff}}(\omega),\nonumber\\
\langle \tilde{b}_{in,\mathrm{eff}}(t)\tilde{b}_{in,\mathrm{eff}}^{\dagger}(t')\rangle &=& \gamma_{\mathrm{eff}}\int \frac{d\omega}{2\pi^2}e^{-i\omega(t-t')}[1+n_{b,\mathrm{eff}}(\omega)],
\eea
with $n_{b,\mathrm{eff}}(\omega)=[\exp(\omega/T_{\mathrm{eff}})-1]^{-1}$. Eq. (\ref{eq:b_effective}) will be adopted to calculate the mechanical correlation function involved in the charge current in the below.

As for the electronic operator, we have
\bea\label{eq:dot_d}
\dot{d}(t) &=& -\left[i\left(\epsilon_0-\frac{\lambda^2}{\omega_b}+2\frac{\lambda g}{\omega_b}n_p\right)+\sum_v\Gamma_v\right]d(t)-i\sqrt{2\pi}\sum_v\mathcal{D}_{\lambda}(t)d_{in}^v(t)\nonumber\\
&=& -\left(i\tilde{\epsilon}_n+\Gamma\right)d(t)-i\sqrt{2\pi}\sum_v\mathcal{D}_{\lambda}(t)d_{in}^v(t),
\eea
here we have defined $\Gamma\equiv\sum_v\Gamma_v$ and
\begin{equation}\label{eq:s13}
\tilde{\epsilon}_n~\equiv~\epsilon_0-\frac{\lambda^2}{\omega_b}+2\frac{\lambda g}{\omega_b}\bar{n}_p.
\end{equation}
We note that the dynamical evolution of displacement operator $\mathcal{D}_{\lambda}(t)$ is now determined by the effective description Eq. (\ref{eq:b_effective}).

The average charge current out of the source in the steady state limit reads
\begin{equation}\label{eq:a1}
 \langle J_S\rangle~=~2\sqrt{2\pi}\mathrm{Im}\left\langle\tilde{d}^{\dagger}d_{in}^S\right\rangle-2\Gamma_S\langle d^{\dagger}d\rangle.
\end{equation}
Here, the ensemble average are evaluated with respect to an initial factorized state where the metallic leads and mechanical thermal environment are in their thermal equilibrium states. To get the explicit form of $\langle J_S\rangle$, we solve Eq. (\ref{eq:dot_d}) in the steady state limit of $t_0\to-\infty$:
\begin{equation}
d(t)~=~-i\sqrt{2\pi}\sum_v\int_{-\infty}^t e^{-(\Gamma+i\tilde{\epsilon}_n)(t-\tau)}\mathcal{D}_{\lambda}(\tau)d_{in}^v(\tau).
\end{equation}

We first evaluate the ensemble average $\left\langle\tilde{d}^{\dagger}d_{in}^S\right\rangle$ on the right-hand-side (RHS) of Eq. (\ref{eq:a1}) by using the correlation functions for input fields listed in Eqs. (\ref{eq:corr_input}) for $d_{in}^v$ and (\ref{eq:bin_corr}) for $b_{in}$:
\bea
\langle \tilde{d}^{\dagger}d_{in}^S\rangle &=& i\sqrt{2\pi}\int_{-\infty}^t\,d\tau e^{-(\Gamma-i\tilde{\epsilon}_n)(t-\tau)}\langle d_{in}^{S,\dagger}(\tau)\mathcal{D}_{\lambda}^{\dagger}(\tau)\mathcal{D}_{\lambda}(t)d_{in}^S(t)\rangle\nonumber\\
&\simeq& i\sqrt{2\pi}\Gamma_S\int\,d\epsilon\frac{n_F^S(\epsilon)}{2\pi^2}\int_0^{\infty}e^{-(\Gamma-i\tilde{\epsilon}_n+i\epsilon)\tau}B_{\lambda}^{\ast}(\tau)d\tau,
\eea
where we have decoupled the electron and mechanical mode correlations by noting that Eq. (\ref{eq:b_effective}) does not contain any electronic operators,
\begin{equation}
\langle d_{in}^{S,\dagger}(\tau)\mathcal{D}_{\lambda}^{\dagger}(\tau)\mathcal{D}_{\lambda}(t)d_{in}^S(t)\rangle~\simeq~\langle d_{in}^{S,\dagger}(\tau)d_{in}^S(t)\rangle\langle\mathcal{D}_{\lambda}^{\dagger}(\tau)\mathcal{D}_{\lambda}(t)\rangle
\end{equation}
and introduced a mechanical mode correlation function $B_{\lambda}(t-\tau)=\langle \mathcal{D}_{\lambda}^{\dagger}(t)\mathcal{D}_{\lambda}(\tau)\rangle$ whose detailed form reads \cite{Liu.20.PRB}:
\begin{eqnarray}
\label{eq:corr_bb}
B_{\lambda}(\tau) &=& \exp\Bigg[-\frac{\lambda^2}{\omega_b^2}\int \frac{d\omega}{\pi}\frac{\gamma_{\mathrm{eff}}}{\gamma_{\mathrm{eff}}^2+(\omega-\omega_{b})^2}\Big(\mathrm{coth}(\omega/2T_{\mathrm{eff}})(1-\cos\omega\tau)+i\sin\omega\tau\Big)\Bigg].
\end{eqnarray}
Similarly, we find
\bea
\langle d^{\dagger}d\rangle &=& 2\int\,d\epsilon\frac{\Gamma_Sn_F^S(\epsilon)+\Gamma_Dn_F^D(\epsilon)}{2\pi}\int_{-\infty}^{t}d\tau\int_{-\infty}^{t}d\tau'e^{i(\epsilon-\tilde{\epsilon}_n)(\tau-\tau')}e^{-\Gamma(2t-\tau-\tau')}B_{\lambda}(\tau-\tau')\nonumber\\
&=& 4\int\,d\epsilon\frac{\Gamma_Sn_F^S(\epsilon)+\Gamma_Dn_F^D(\epsilon)}{2\pi}\mathrm{Re}\Big[\int_{-\infty}^{t}d\tau\int_{-\infty}^{\tau}d\tau'e^{i(\epsilon-\tilde{\epsilon}_n)(\tau-\tau')}e^{-\Gamma(2t-\tau-\tau')}B_{\lambda}(\tau-\tau')\Big]\nonumber\\
&=&2 \int\,d\epsilon\frac{\Gamma_Sn_F^S(\epsilon)+\Gamma_Dn_F^D(\epsilon)}{2\pi\Gamma}\mathrm{Re}\Big[\int_{0}^{\infty}d\tau e^{-(\Gamma+i\tilde{\epsilon}_n-i\epsilon)\tau}B_{\lambda}(\tau)\Big].
\eea
Here, ``Re" takes the real part. Altogether, we find
\begin{equation}\label{eq:e7}
\langle J_S\rangle~=~\frac{4\Gamma_S\Gamma_D}{\Gamma}\int\,\frac{d\epsilon}{2\pi}\mathrm{Re}\Big[\int_{0}^{\infty}d\tau e^{-(\Gamma+i\tilde{\epsilon}_n-i\epsilon)\tau}B_{\lambda}(\tau)\Big][n_F^S(\epsilon)-n_F^D(\epsilon)],
\end{equation}
which is the charge current expression that we use in the main text. Notably, we can identify an effective transmission function in the integral. It depends on the mechanical mode autocorrelation function, and it includes the backaction of electrons through an effective-bath approximation.

\section{II. Measuring photon number with an asymmetric bias drop}
\label{a:2}
In general, we can express the chemical potentials of the electrodes (source and drain) as 
\begin{equation}\label{eq:mu_split}
\mu_S~=~\alpha V,~~\mu_D~=~-(1-\alpha)V.
\end{equation}
Here $V$ is the voltage bias across the SET and $\alpha\in$ [0, 1] characterizes the asymmetry of bias drop and can be determined by experiments. For a solid-state SET, we can have a phenomenological expression
\begin{equation}
    \alpha~=~\frac{R_S}{R_S+R_D}
\end{equation}
with $R_v$ ($v=S, D$) the junction resistance of the corresponding island-lead interface. For a nanotube-based SET, we have $\alpha=\Gamma_D/(\Gamma_S+\Gamma_D)$ instead. If $R_S=R_D$ or $\Gamma_S=\Gamma_D$, we recover the symmetric bias drop considered in the main text.

With the above voltage splitting given by Eq. (\ref{eq:mu_split}), the resonant electron transport occurs when the following condition
\begin{equation}\label{eq:s23}
    \mu_S~=~\tilde{\epsilon}_n
\end{equation}
is fulfilled. Here, $\tilde{\epsilon}_n$ is given by Eq. (\ref{eq:s13}). From the above equation, we find
\begin{equation}
    V_n^{\ast}~=~\frac{\tilde{\epsilon}_n}{\alpha},
\end{equation}
which yields the following expression for the measured photon number
\bea
    \bar{n}_{p,\mathrm{measure}} &=& \frac{\alpha}{\Delta\epsilon}(V_n^{\ast}-V_0^{\ast}).
\eea
Here, $\Delta\epsilon\equiv2\lambda g_0/\omega_b$. The symmetric case considered in the main text is recovered when $\alpha=1$.

\end{widetext}

\end{document}